\documentclass[authoryear]{elsarticle}

\renewcommand{\vec}[1]{{\bf #1}}
\usepackage{color, graphicx}

\journal{Advances in Water Resources}
\begin{document}
\begin{frontmatter}

\title{Viscous and Gravitational Fingering in Multiphase Compositional and Compressible Flow}
\author{Joachim Moortgat}
\address{School of Earth Sciences, The Ohio State University, 275 Mendenhall Laboratory, 125 S.~Oval Mall, OH 43210, USA.\\moortgat.1@osu.edu\footnote{\copyright~2016. This manuscript version is made available under the CC-BY-NC-ND 4.0 license http://creativecommons.org/licenses/by-nc-nd/4.0/ }}

\begin{abstract}
Viscous and gravitational fingering refer to flow instabilities in porous media that are triggered by adverse mobility or density ratios, respectively. These instabilities have been studied extensively in the past for 1) {single-phase} flow (e.g., contaminant transport in groundwater, first-contact-miscible displacement of oil by gas in hydrocarbon production), and 2) multi-phase {immiscible} and incompressible flow (e.g., water-alternating-gas (WAG) injection in oil reservoirs). Fingering in \textit{multiphase compositional} and compressible flow has received much less attention, perhaps due to its high computational complexity. However, many important subsurface processes involve multiple phases that exchange species. Examples are carbon sequestration in saline aquifers and enhanced oil recovery (EOR) by gas or WAG injection below the minimum miscibility pressure. In multiphase flow, relative permeabilities affect the \textit{mobility} contrast for a given  {viscosity} ratio. Phase behavior can also change local fluid properties, which can either enhance or mitigate viscous and gravitational instabilities. This work presents a detailed study of fingering behavior in compositional multiphase flow in two and three dimensions and considers the effects of 1) Fickian diffusion, 2) mechanical dispersion, 3) flow rates, 4) domain size and geometry, 5) formation heterogeneities, 6) gravity, and 7) relative permeabilities.  
Results show that fingering in compositional multiphase flow is profoundly different from miscible conditions and upscaling techniques used for the latter case are unlikely to be generalizable to the former. 
\end{abstract}

\begin{keyword}
viscous fingering\sep multiphase compositional flow\sep phase behavior\sep flow instabilities \sep simulation \sep wettability
\PACS 47.11.Fg\sep 47.20.-k \sep 47.20.Bp \sep 47.20.Gv \sep47.56.+r
\end{keyword}

\end{frontmatter}

\section{Introduction}
Gravitational and viscous flow instabilities can occur both within a single phase or when multiple fluid phases flow through the same porous media. 
A few examples for single-phase flow are 1) the spreading of a contaminant or solvent that changes the density or viscosity of an aqueous phase upon dissolution \citep{tan1986stability, schincariol1997instabilities}, 2) enhanced oil recovery (EOR) by first-contact-miscible (FCM) gas injection \citep{moissis1993simulation}, and 3) carbon sequestration in saline aquifers \citep{pruess3}. The latter only considers the local density increase in an aqueous phase upon CO$_{2}$ dissolution, which can trigger gravitational fingering throughout the aquifer. 

Injection of low-viscosity, high-density water into a reservoir saturated with lighter but more viscous oil, is an example where both viscous and gravitational instabilities may occur for two-phase immiscible, and often incompressible flow. Migration of dense-non-aqueous-phase liquids (DNAPL) through groundwater is in a sense the opposite problem. 
Studies of water-alternating-gas (WAG) injection also often assume two-phase flow: the gas is FCM in the oil, while the aqueous phase is immiscible. 

The most complicated processes susceptible to fingering involve multiphase compositional and compressible flow. Important examples are: 1) EOR by gas injection \textit{below} the minimum miscibility pressure (MMP), 2) WAG below the MMP for the injected gas, 3) carbon sequestration, taking into account the CO$_{2}$-rich gas phase. Another example is injection of CO$_{2}$ on top of denser oil. This should be gravitationally stable, but when CO$_{2}$ \textit{dissolves} it can increase the oil density in the top. This is unstable to gravitational fingering \textit{within} the oil phase \citep{ahmed2012complex, shahraeeni2015high}, similar to the driver of fingering in carbon sequestration.  

It is hard to do justice to all important contributions in the vast literature on fingering behavior in porous media. The following review is intended to put this work into the context of earlier studies, which were mostly confined to single-phase flow. 

Most studies were carried out in the 1980s and '90s for \textit{miscible} (FCM) displacement, motivated by earlier 
Hele-Shaw experiments (e.g., \citet{hill1952channeling, chuoke1959instability, saffman1958penetration, benham1963model}). \citet{todd1972development} proposed a correlated upscaling technique, which has been widely used in commercial reservoir simulators to mimic the effect of small-scale fingering behavior on coarse grids. 
\citet{tan1986stability, tan1987stability,tan1988simulation} performed experiments, linear stability analyses, and some of the earliest numerical simulations of the non-linear instability regime (reviewed in \citet{homsy1987viscous}). \citet{zimmerman1992viscous, zimmerman1991nonlinear} also considered the effects of anisotropic (mechanical) dispersion, while the effects of formation heterogeneities were investigated by \citet{araktingi1993viscous, tchelepi1994interaction, tan1992viscous,tchelepi1993dispersion,tchelepi1994viscous}.
\citet{moissis1988simulation, moissis1993simulation} presented the state-of-the-art in numerical simulations at that time. 

The above studies were mostly for two-dimensional (2D) flow. Early simulations of fingering in three dimensions (3D) were presented by \citet{zimmerman1992three, christie19933d, tchelepi1994viscous}. \textit{Gravitational} fingering, or density driven flow, impacting single-phase solute transport in groundwater was investigated by, among others, \citet{shikaze1998density, schincariol1997instabilities,zhang1995multispecies,schincariol1990experimental}, using both experiments and numerical simulations. 

\citet{blunt1994theory} considered fingering in two-phase three-component flow. A solvent is still FCM in oil, but an immiscible aqueous phase is considered as well, and both phases are assumed incompressible. These assumptions form the basis for most studies of WAG injection to date (e.g., \citet{juanes2006impact}). \citet{blunt1994predictive} generalized the \citet{todd1972development} model to two-phase flow. 
A few more recent studies presented experiments of heavy oil displacement by solvent \citep{cuthiell2006viscous}, higher-order finite element simulations of viscous fingering in single-phase flow 
\citep{scovazzi2013discontinuous, huang2013high, scovazzi2013,gerstenberger2013computing}, and experiments and stability analyses for forced imbibition \citep{sharma2012experiments}. 

Carbon sequestration in saline aquifers is one important application where \textit{gravitational} fingering may be critical, particularly when CO$_2$ has accumulated in the top of the aquifer. When CO$_{2}$ dissolves into the brine, it can cause a small increase of the aqueous phase density in the top \citep{garcia2001density, duan2008densities}. This can trigger gravitational fingering, which effectively mixes dissolved CO$_{2}$ throughout the aquifer, because the convective time-scales for high permeability formations are much shorter than for diffusive transport. The literature on this process is extensive and will not be reviewed in detail here (see, e.g., \citet{ennis2003role,xu06,riaz, pruesszhang,pruess3, philipsequestration} and references therein). From a modeling perspective, the problem is similar to FCM flow of a solvent in a weakly compressible fluid.

This short literature review illustrates that both viscous and gravitational flow instabilities have been studied in great detail, through experimental, analytical (stability analyses) and numerical investigations. However, all the aforementioned studies assume that the adverse viscosity and density 
contrasts are caused by a solvent that is fully dissolved (FCM) in the displaced fluid, sometimes also considering an immiscible and incompressible second, aqueous, phase. Two studies of fingering in multiphase compositional flow were carried out by \citet{blunt1994predictive,chang1994co2}. However, limitations in computational power at that time only allowed for simulations on relatively coarse grids, even on a Cray system. 
\citet{chang1994co2} found that fingering behavior in compositional multiphase flow is different from FCM displacement, due to relative permeability and phase behavior. The authors acknowledged that more detailed simulations are required on finer grids. 

The objective of this study is to do just that: to investigate fingering in multiphase flow with considerable mass transfer between the phases on fine grids, taking advantage of increased computational power and advanced higher-order finite element methods. 
Three fully compositional multicomponent phases are considered: water, oil, and gas. The phase compositions and phase properties are derived from rigorous equation-of-state (EOS) based phase-stability analyses and phase-split computations. Hydrocarbon phases are modeled with the \citet{preos} EOS, and the aqueous phase with the cubic-plus-association (CPA) EOS \citep{liCPA}. 
Viscosities are determined by the \citet{viscosity2} model. All relevant physical processes are taken into account: gravity, anisotropic mechanical dispersion, and Fickian diffusion. The latter is represented by a unique model for multicomponent multiphase flow \citep{kassem, LeahyDios,moortgatVII}. 

Fingering behavior is expected to be different from single-phase flow because of 1) the effect of relative permeabilities, which changes the \textit{mobility} contrast between two phases for a given adverse \textit{viscosity} ratio, and 2) phase behavior effects, particularly local changes in densities and viscosities, which can either enhance or stabilize flow instabilities. The focus of this work is on applications where both of these effects are most pronounced: EOR by gas injection below the MMP, with or without the presence of an aqueous phase (e.g., in WAG). 

{\color{black} In this work, a reservoir oil is considered, which upon mixing with injected CO$_{2}$ (at a given reservoir temperature and pressure) is near the critical point and exhibits significant species exchange and non-trivial phase behavior. \citet{moortgatIV, shahraeeni2015high} were able the model the detailed results of experiments with gravitational fingering at the core scale, including Fickian diffusion but without mechanical dispersion. A single example of viscous fingering during WAG injection in this oil was presented in \citet{moortgatIII}. Results were compared to a commercial reservoir simulator, demonstrating that lowest-order numerical methods cannot resolved the fingers on feasible grid sizes due to numerical dispersion. By using higher-order FE methods the process can be captured on coarse grids suitable for large-scale domains. The aforementioned numerical issues are not revisited here. Instead the focus is on a range of physical processes that affect the character of fingering instabilities.}

The sections that follow include a summary of the main governing equations for multicomponent multiphase compositional flow, a discussion of simulation results, and the key conclusions. 
The analyses themselves consider 1) the importance of anisotropic dispersion and Fickian diffusion as potential restoring forces, 2) the interplay between viscous and gravitational fingering, 3) effects of dimensionality, 4) permeability heterogeneities, and 5) rate and domain size dependencies. The assumptions for this study are: 1) high P\'eclet numbers (advection dominated flow), 2) mobility ratios that are unstable to viscous fingering, and 3) negligible initial gas-oil density contrast, such that gravitational effects are only due to local changes in density from phase behavior.

\section{Problem Set-Up}
Multiphase compositional flow in porous media is described by mass conservation (or transport) equations for each species $i$ {\color{black} (or $j$) in a $n_{c}$-component mixture}, Darcy velocities for each phase $\alpha$ (with $\alpha = g, o, w$ for gas, oil and water phases, respectively), and a pressure equation that involves the formation and fluid compressibilities $C_{r}$, and $C_{f}$. 

\subsection{Advection-Diffusion-Dispersion Transport}
The transport equations (molar balance) are given by
\begin{eqnarray}\label{sec::masscons}
&&\phi\frac{\partial c z_i}{\partial t} + \nabla \cdot \vec{U}_i = F_i, \quad i = 1, \ldots, n_c 
\end{eqnarray}
in terms of porosity, $\phi$, molar density of the multiphase mixture, $c$, overall molar composition, $z_{i}$, and sink and source terms $F_{i}$ that can represent production and injection wells. The divergence term includes advective, diffusive, and dispersive phase fluxes:
\begin{eqnarray}\label{eq::convdiff}
&&\vec{U}_i = \sum_{\alpha} (c_\alpha x_{\alpha,i} \vec{u}_{\alpha} + \vec{J}^{\mathrm{diff}}_{\alpha,i } + \vec{J}^{\mathrm{disp}}_{\alpha,i }),\quad i=1,\ \dots ,\ n_c,\\\label{eq::difff}
&& \vec{J}_{\alpha,i}^{\mathrm{diff}} = - \phi S_\alpha c_\alpha \sum_{j=1}^{n_c -1} D^\mathrm{Fick}_{\alpha, ij}\ \nabla x_{\alpha, j} =-\frac{\phi S_\alpha c_\alpha}{R T} \sum_{j=1}^{n_{c}-1} {\cal B}^{\mathrm{M}}_{\alpha, ij}\ x_{\alpha, j}\nabla \mu_{\alpha, j}, 
\\\label{eq::disp1}
&& \vec{J}_{\alpha,i}^{\mathrm{disp}} = - \phi S_\alpha c_\alpha \sum_{j=1}^{n_c -1} \vec{D}^\mathrm{disp}_{\alpha}\ \nabla x_{\alpha, j},
\end{eqnarray}
with $\vec{u}_{\alpha}$ the Darcy phase velocities (defined in Eq.~\ref{eq::darcy3x}), $S_{\alpha}$ the saturations, $x_{\alpha, i}$ the phase compositions, $c_{\alpha}$ the phase molar density, $T$ the temperature, and $R$ the gas constant. Fickian diffusion can be expressed with either compositional or chemical potential gradients as the driving force (Eq.~\ref{eq::difff}). {\color{black}The latter is more robust in heterogeneous or fractured media \citep{moortgatVII}. Each formulation requires a full matrix of composition dependent coefficients ($D^\mathrm{Fick}_{\alpha, ij}$ and ${\cal B}^{\mathrm{M}}_{\alpha, ij}$)}.

Mechanical dispersion is governed by the anisotropic tensor:
\begin{eqnarray}\label{eq::disp2}
&&\vec{D}^\mathrm{disp}_{\alpha} = d_{t,\alpha} | \vec{u}_{\alpha} | \vec{I}  + (d_{l,\alpha} - d_{t,\alpha}) \frac{\vec{u}_{\alpha} \vec{u}^{T}_{\alpha}}{|\vec{u}_{\alpha}|}
\end{eqnarray}
with $d_{l,\alpha}$ and $d_{t,\alpha}$ the longitudinal and transverse dispersivities, respectively, $\vec{I}$ the identity matrix, and $|\vec{u}_{\alpha}|$ the magnitude of the phase velocity vector. Diffusion and dispersion may also be affected by tortuosity.

\subsection{Darcy Flow}\label{sec::darcy}
The Darcy phase velocities are given by
\begin{eqnarray}\label{eq::darcy3x}
&&{{\mathbf u}}_{\alpha }= - \lambda_{\alpha} {\mathrm K} (\nabla p_\alpha  -{\rho }_{\alpha } {\mathbf g}),
\end{eqnarray}
where the phase pressures $p_{\alpha}$ are generally expressed in terms of a reference pressure (oil) and two capillary pressures $p_{c,go} =p_{g} - p_{o}$ and 
$p_{c,wo} =p_{o}- p_{w}$ that are functions of phase saturations \citep{moortgatV}. {\color{black}Capillary effects will not be considered in this work, which studies fluid at low interfacial tension conditions.}
Differences in phase mass densities, $\rho_{\alpha}$, are the drivers for gravitational fingering. 

An important observation is that in \textit{multiphase} flow, the phase velocities are proportional to \textit{mobilities} $\lambda_{\alpha} = k_{r,\alpha}/\mu_{\alpha}$ that depend not only on the phase viscosities $\mu_{\alpha}$, but also on the relative permeabilities $k_{r,\alpha}$. {\color{black}Consider water-oil flow with an adverse viscosity ratio of $10$ and Corey relative permeabilities without residual saturations and with an end-point relative permeability of $0.4$ for water (different relations are used in the Numerical Experiments):}
\begin{eqnarray}
&& k_{r,w} = 0.4 S_{w}^{3}\quad\mbox{and}\quad k_{r,o} = (1-S_{w})^{2}.
 \end{eqnarray}
The mobility ratio in a two-phase region (i.e., not across a phase front) then scales as 
 \begin{eqnarray}
 M = \frac{\lambda_{w}}{\lambda_{o}} = \frac{4 S_{w}^{3}}{(1-S_{w})^{2}},
  \end{eqnarray}
which is much less than one for $S_{w}<  60\%$, and much larger than 10 for high $S_{w}$. The same is true for linear relative permeabilities with unit end-points ($M = 10 S_{w}/(1-S_{w})$). 
  
  The implication is that while flow instabilities in single-phase \textit{miscible} displacement are mostly affected by fluid properties (viscosity and density), rock wettability plays a critical role in fingering for multiphase flow with important consequences for EOR. These effects have not been explored in detail in the fingering literature.
  
  \subsection{Pressure equation for compressible flow}
The pressure equation is derived by \citep{acs,watts} from volume balance, and is given for $p_{o}$ as:
 \begin{eqnarray} \label{GrindEQ__3_} 
&&\phi (C_{r}+C_{f}) \frac{\partial p_o}{\partial t} + \sum_{i=1}^{n_c} {\overline{\nu}_i(\nabla \cdot \vec{U}_i}-F_i) =0. 
\end{eqnarray}
Eq.~\ref{GrindEQ__3_} accounts for the compressibilities of the rock and all three fluid phases \citep{moortgatIII}. Partial molar volumes in the mixture are denoted by $\overline{\nu}_i$.

Phase compositions ($x_{\alpha, i}$) and the molar fractions of each phase are obtained from phase-split computations, and the mathematical framework is closed with constraints ($\sum_{i}z_{i} = \sum_{i} x_{\alpha,i}=1$, etc.). Overlapping Dirichlet and Neumann boundary conditions can accommodate constant injection rates and production at constant rate or pressure.

To capture the small-scale onset of viscous and gravitational fingers a higher-order discontinuous Galerkin (DG) method is used to update the transport equations. A Mixed Hybrid Finite Element (MHFE) method provides accurate velocities, particularly for heterogeneous and anisotropic permeability fields (and on unstructured grids). Details of the numerical methods are presented in earlier work \citep{moortgatI,moortgatVI, moortgatV,moortgatII, moortgatIII, moortgatIV,shahraeeni2015high}.
The key point is that these methods have low numerical dispersion and can resolve fine-scale fingers on coarse grids{\color{black}, due to the higher convergence rate to the `true' solution, which makes the approach highly computationally efficient.} The numerical examples in this work involve millions of degrees-of-freedom (DOF) but are carried out in serial on a personal computer. Such simulations were not feasible at the time when much of the earlier research on fingering behavior took place.

\section{Numerical Experiments}
To best illustrate the differences in fingering behavior between miscible displacement and multiphase compositional flow, a 9 (pseudo-)component oil is considered that has been used in several earlier papers as a benchmark of challenging compositional modeling. The fluid composition, critical properties, and a detailed PVT analysis are provided in (Table~\ref{table::fluid} and \citep{moortgatIV}). The reservoir temperature is $T=58^{\circ}$C and the initial pressure at the bottom is 441 bar.

All the simulations are for lateral displacement of oil by a (80/20 mol\%)  CO$_2$/methane gas mixture in a rectangular (2D or 3D) domain. At the initial conditions the oil and injection gas densities are $\rho_{o}=736\ \mathrm{kg}/\mathrm{m}^{3}$ and $\rho_{g} = 731\ \mathrm{kg}/\mathrm{m}^{3}$ and the viscosities are $\mu_{o}= 1.28$ cp and $\mu_{g}=0.06$ cp (Table~\ref{table::fluid}).
Significant viscous fingering occurs because, although the gas has almost the same density as the oil at the reservoir conditions, the viscosity ratio of $\mu_{o}/\mu_{g}=21$ is adverse. If the two phases were immiscible, one would expect viscous fingering to be pronounced but gravity effects to be small. Other sources of complexity can exist as well. For example, Section~\ref{sec::darcy} discussed how relative permeabilities may affect fingering instabilities in multiphase flow. \textit{Phase behavior} is the most critical source of additional complexity in \textit{compositional} multiphase flow. Consider a single phase-split calculation for one mole of the initial oil mixed with two moles of the injection gas. For this mixture the phase densities are $\rho_{o}=818\ \mathrm{kg}/\mathrm{m}^{2}$ and $\rho_{g} = 647\ \mathrm{kg}/\mathrm{m}^{2}$ and the viscosities are $\mu_{o}= 1.09$ cp and $\mu_{g}=0.09$ cp. In other words, the viscosity contrast is reduced to $\mu_{o}/\mu_{g}=12$ while the density contrast becomes $(\rho_{o}-\rho_{g})/\rho_{o} = 10\%$. This illustrative example suggests that compositional effects may reduce the degree of viscous fingering, but at the same time may trigger gravitational fingering and/or gravity override. As will be apparent in the examples, both types of fingers can occur.

Simulations are carried out on a $600\ \mathrm{m}$ wide and $60\ \mathrm{m}$ high domain. {\color{black} For two-dimensional simulations, we will sometimes refer to a \textit{horizontal} cross-section to indicate that gravity is neglected, whereas \textit{vertical} domains include gravity. Additionally, we use the common notation of $y$ as a horizontal direction and $z$ always refers to a vertical axis.}
The aspect ratio of $L_{x}/L_{z}=10$ favors viscous over gravitational effects. Gas is injected uniformly along the left boundary at a constant rate. Production is from the right boundary at a constant pressure. The grid blocks are 1 m in each direction, which is sufficiently fine for higher-order FE results to converge. The porosity is 13\% and the permeability field is a lognormal random distribution with a standard deviation of $2.5\%$ around a mean of 133 md. This variation is sufficient to trigger flow instabilities without affecting their subsequent evolution.
Because of the high reservoir pressure and strong species exchange (mass transfer between the phases), capillary pressures are negligible and Brooks-Corey relative permeabilities can be assumed linear with unit end-points. The residual oil saturation is 10\%, but the oil saturation can easily fall below the residual value due to evaporation. 

\subsection{Base Case}
\subsubsection{Without Gravity}
\begin{figure}[!h]
\centerline{\includegraphics[width=\textwidth]{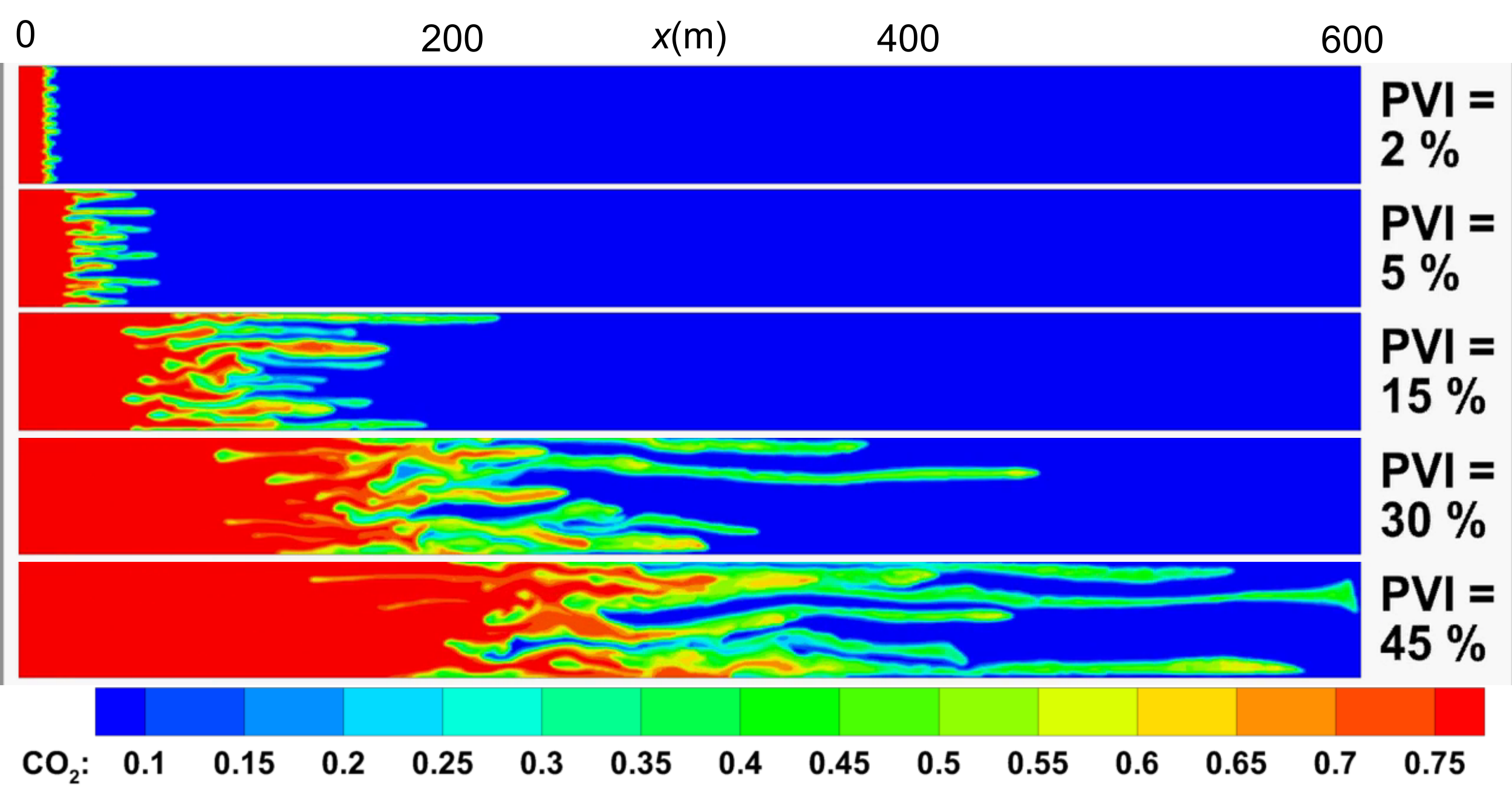}}
\caption{\label{fig2}Overall CO$_{2}$ molar fraction for gas injection in horizontal domain at $7.3\%$ PV/yr without diffusion or dispersion. (Legend for CO$_{2}$ composition is the same in all following similar figures.)}
\end{figure}
The first example is a simplified set-up to serve as a base case for comparison. Diffusion and dispersion are neglected and the domain is a horizontal 2D cross-section (no gravity), such that viscous fingers do no interfere with gravitational ones. The gas (CO$_{2}$ and methane) injection rate is $7.3\%$ pore volume (PV) per year. Figure~\ref{fig2} shows the overall CO$_{2}$ composition throughout the domain at different times (the color scale for CO$_{2}$ is the same in all subsequent figures). 

About 15 small-scale viscous fingers develop almost instantly (clearly visible at 2\% pore volume injected, or PVI), corresponding to a characteristic wavelength of 4 m. At later times the fingers grow, split, shield each other, and merge. Only a few larger fingers remain around the time of breakthrough (45\% PVI). Apart from these qualitative aspects, fingering is different from miscible displacement due to relative permeabilities: {\color{black}once gas saturation increases locally with respect to a neighboring region (e.g., at the initial growth of a finger) the effective gas permeability, $k_{r,g}(S_g)\times\mathrm{K}$ also increase.} This feature may amplify the fingering flow, equivalent to channeling in connected high (absolute) permeability regions. 

\begin{figure}[!h]
\centerline{\includegraphics[width=\textwidth]{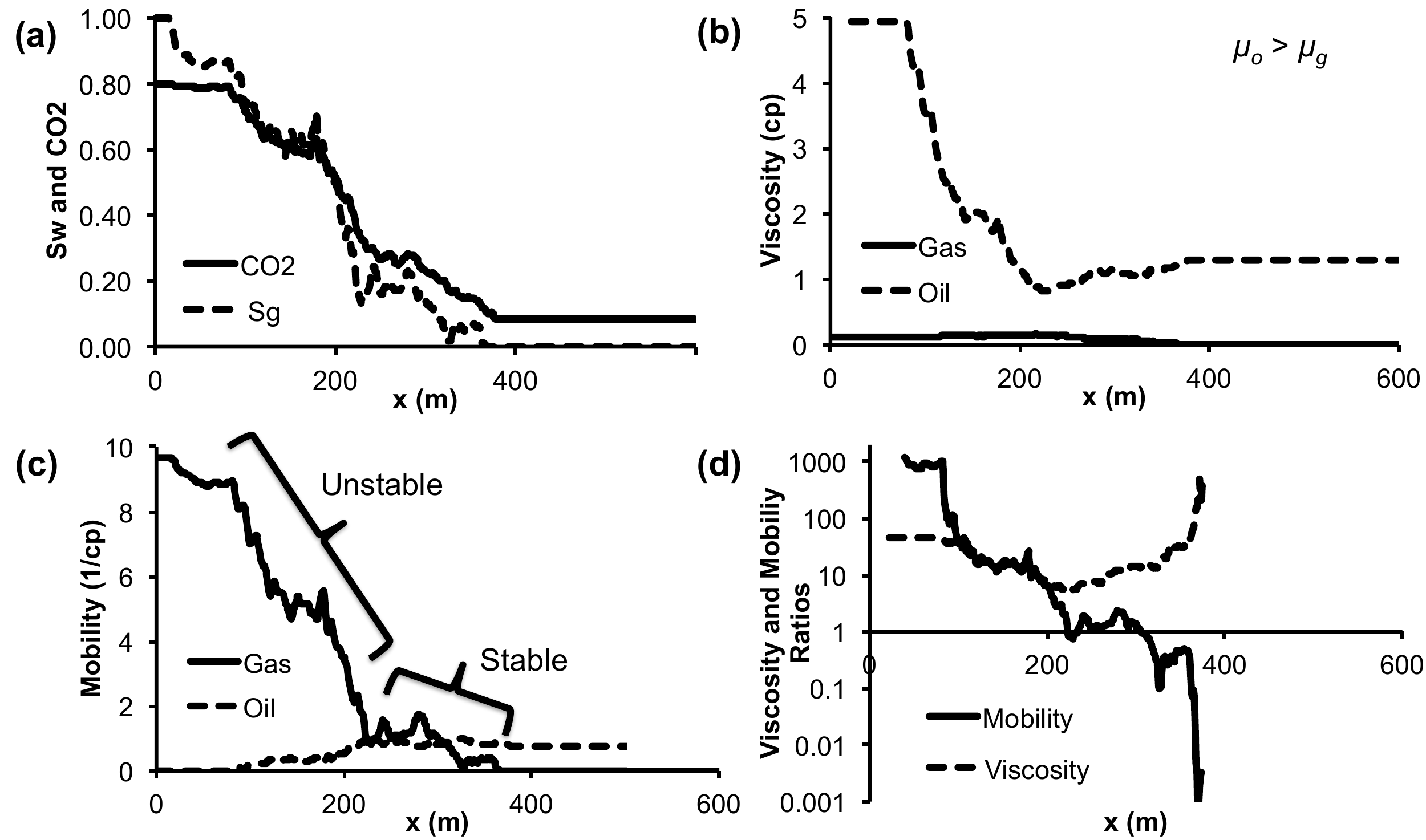}}
\caption{\label{fig3}Gas saturation and overall CO$_{2}$ molar fraction \textbf{(a)}, oil and gas viscosity \textbf{(b)}, oil and gas mobilities \textbf{(c)}, mobility and viscosity ratios \textbf{(d)}, all averaged over the $y$-direction at 30\% PVI.}
\end{figure}

\citet{todd1972development,blunt1994predictive}, and others, have shown that for miscible flow one can predict the average `spreading' 
of a solvent front due to fingering when integrated over the direction normal to flow. This technique has been used in commercial reservoir simulators to make some accommodation for fingering in coarse grid simulations that cannot resolved the fingers. An obvious question is whether this approach could be generalized to multiphase compositional flow.

{\color{black}To investigate this, we similarly integrate the gas saturation, overall CO$_{2}$ composition, phase viscosities and mobilities over the $y$-direction in Figure~\ref{fig3} (at 30\% PVI).} The oil viscosity changes considerably due to species transfer. At low gas saturations, CO$_{2}$ dissolves in oil and decreases the oil viscosity, but at higher saturations (and a constant feed), light oil components {\color{black}(such as methane)} evaporate and are carried away by the gas, leaving behind a more dense and viscous oil. Figure~\ref{fig3} shows that the average viscosity ratio $\mu_{o}/\mu_{g}$ is increased throughout much of the domain compared to the initial viscosity ratio of $21$.

The average phase \textit{mobilities}, which take into account relative permeability and determine the phase velocities, {\color{black}provide a different perspective}. From the mobilities and their ratio (Figure~\ref{fig3}), there appears to be a large region {\color{black}(between 200 and 375 m)} where the gas and oil mobilities become comparable. This would suggest a stabilizing effect, but is instead an artifact of the integration over the transverse direction. Relative permeabilities, unlike viscosities, are functions of saturations. From the 30\% PVI panel in Figure~\ref{fig2} it is clear that the \textit{average} gas mobility is only low for $x>250$ m because there are fewer fingers in that region. The \textit{individual} fingers are highly unstable with large mobility contrasts with respect to oil.

These observations suggest that it might be hard to conceive a \citet{todd1972development} type upscaling procedure for multiphase compositional flow. If the fingering region had a well-defined leading and trailing front with reasonably uniform fingering in between (as in \citet{blunt1994predictive}), one could construct some type of average saturation profile. But when a small number of skinny fingers significantly outpace the others such a procedure becomes less obvious. This issue is reminiscent of permeability upscaling: it is relatively straightforward to upscale single-phase flow through parallel layers with moderate variations in permeability, but harder to do so for two- phase flow when one of the layers has a much higher permeability than the others.

\subsubsection{With Gravity}
\begin{figure}[!h]
\centerline{\includegraphics[width=\textwidth]{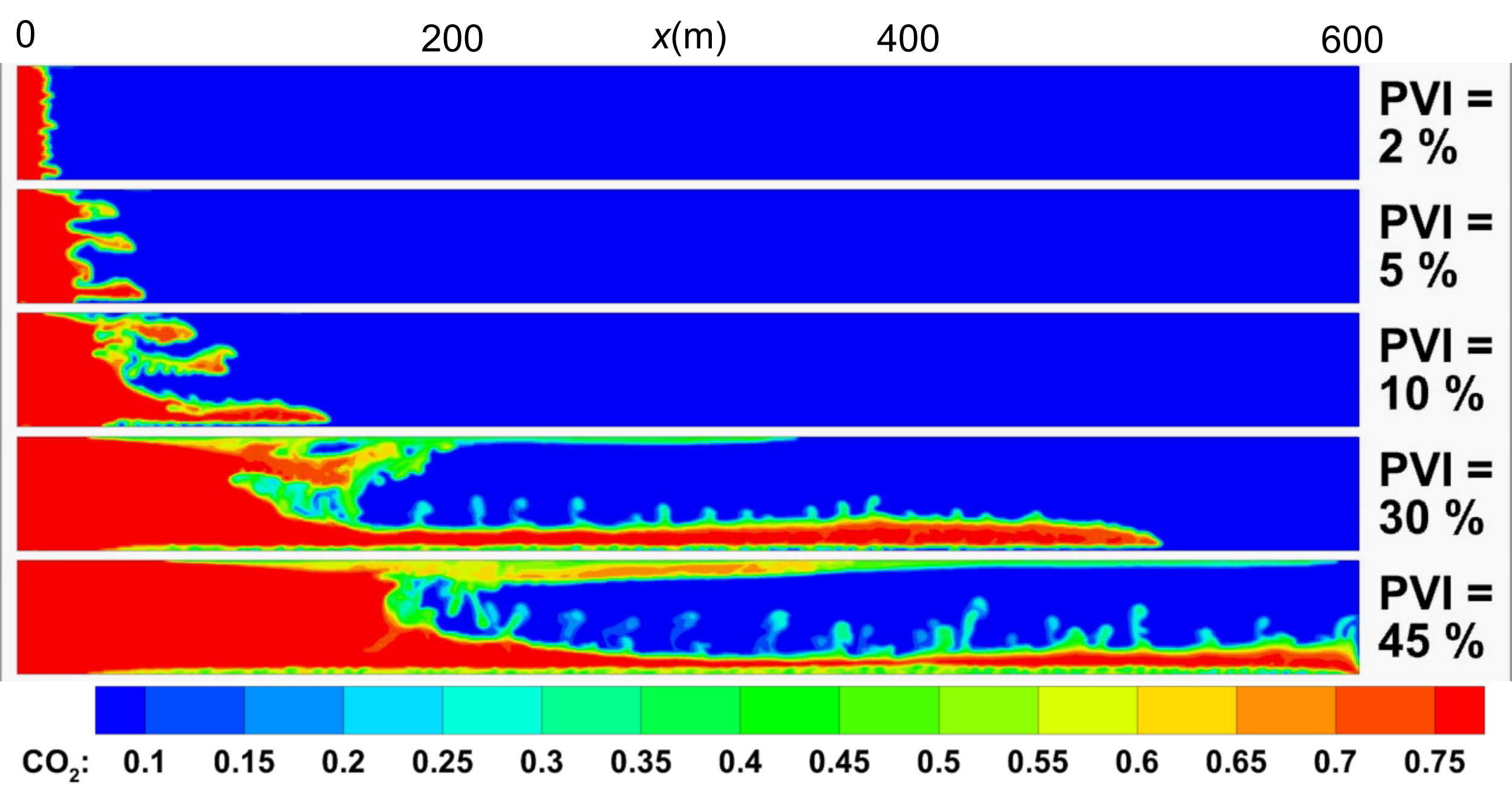}}
\caption{\label{fig4}Overall CO$_{2}$ composition for gas injection in vertical domain at $7.3\%$ PV/yr but without diffusion or dispersion.}
\end{figure}
\begin{figure}[!h]
\centerline{\includegraphics[width=\textwidth]{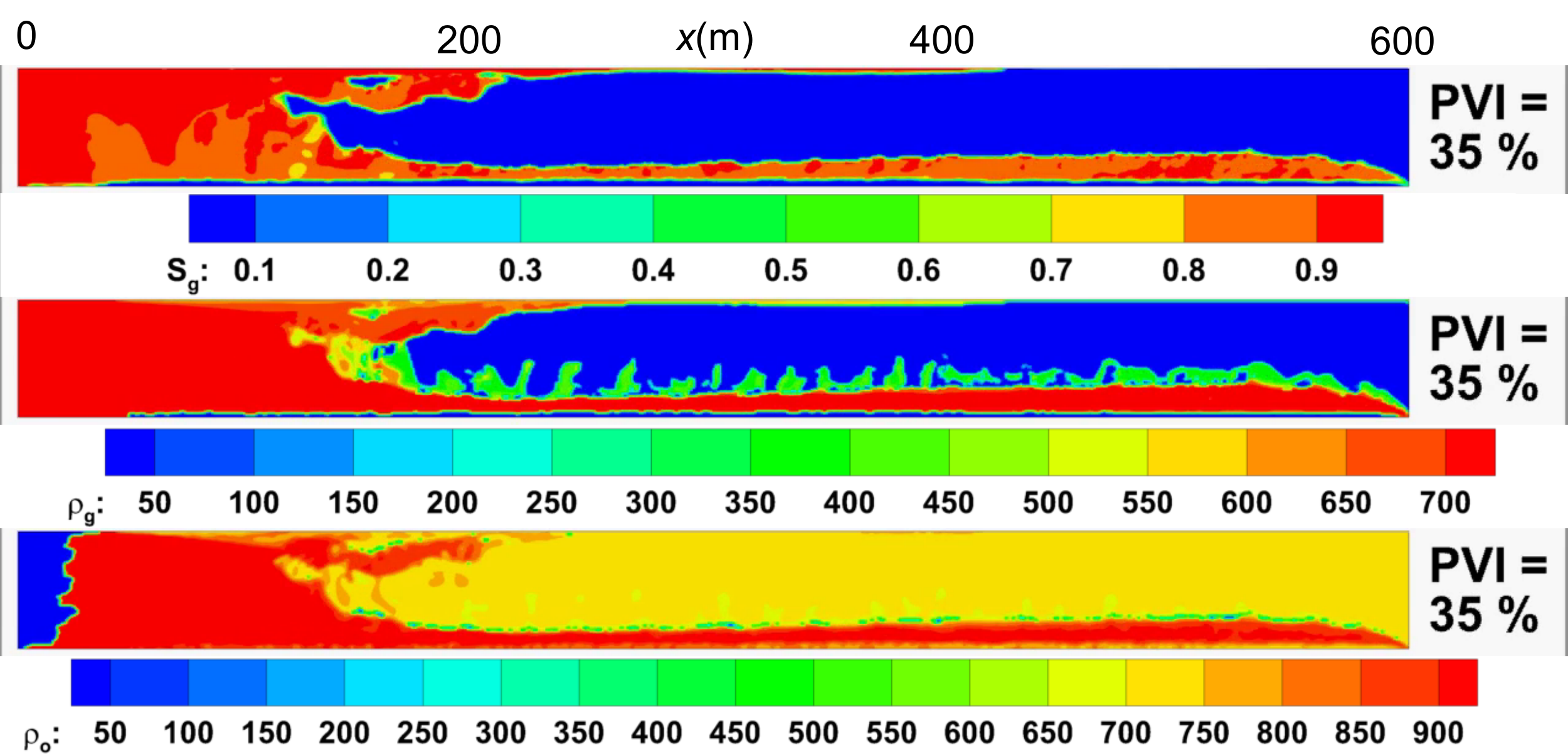}}
\caption{\label{fig5}Gas saturation, gas and oil density (kg/m$^{3}$), for gas injection in vertical domain at $7.3\%$ PV/yr but without diffusion and dispersion.}
\end{figure}

Gravity is added to the base case simulation to consider its effect on viscous fingering.
Remember that the injection gas composition was chosen to have approximately the same density as the oil in place to prevent gravity override {\color{black}(Table~\ref{table::fluid})}. Figure~\ref{fig4} clearly shows, however, that gravity has a significant effect. This is purely due to compositional phase behavior affecting local phase densities throughout the domain (Figure~\ref{fig5}). Lighter gas, enriched by methane evaporating from the oil phase, accumulates at the top of the domain and small gravitational fingers propagate both {\color{black}upwards (from the bottom between $x = 200$ and 600 m) and downwards (before $x \sim 175$ m) out of larger-scale viscous fingers.} The interference of this vertical flow component with horizontal viscous fingers at intermediate heights has a stabilizing effect. The gas under-ride at the bottom is not due to gravity segregation (discussed further in Section~\ref{sec::endpoints}) but to a viscous finger that was able to grow without disturbances by gravitational plumes from below. 
Figure~\ref{fig5} shows the gas saturation and phase densities and indicates that the bottom finger has a gas density close to the oil above it.
The gravitational fingers are due to highly non-linear phase behavior: methane evaporation from oil decreases the gas density and increases the oil density, while elsewhere methane and CO$_{2}$ dissolution decrease the local oil density. Together, the gravitational effects result in breakthrough occurring at 35\% PVI, or 22\% earlier than without gravity (Figure~\ref{fig2}). To predict this type of behavior when gas is injected below the MMP requires detailed composition simulations.  

\subsection{Effects of Mechanical and Fickian Dispersion}\label{sec::diff}
\begin{figure}[!h]
\centerline{\includegraphics[width=\textwidth]{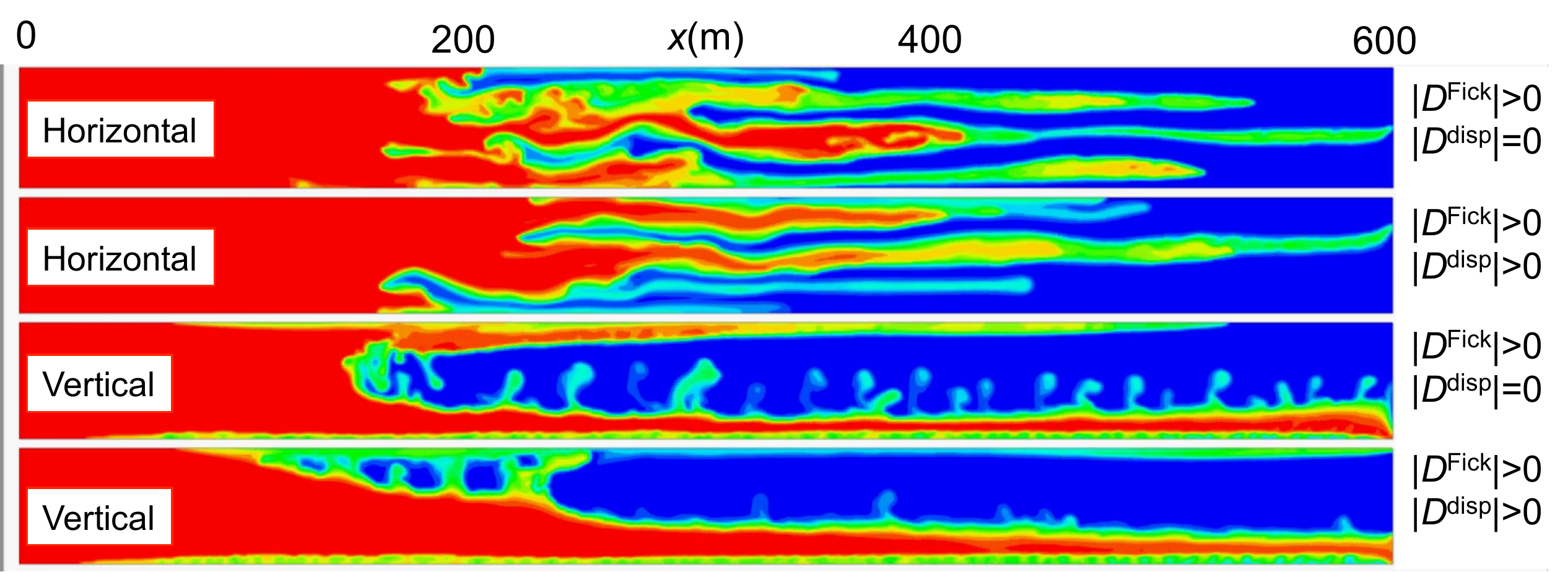}}
\caption{\label{fig6}Overall CO$_{2}$ composition as in Figure~\ref{fig2} but for simulations with Fickian diffusion ($|D^{\mathrm{Fick}}|>0$) and without/with mechanical dispersion ($|D^{\mathrm{disp}}|>0$), for horizontal and vertical 2D cross-sections at 45\% PVI.}
\end{figure}
To model multicomponent multiphase flow more comprehensively, anisotropic mechanical dispersion and Fickian diffusion are included in the next set of simulations. Mechanical dispersion based on Eqs~\ref{eq::disp1}-\ref{eq::disp2} was implemented into the simulator for this work. 
Mechanical dispersion essentially mimics unresolved sub-grid-size heterogeneity, so dispersive length-scales are smaller than the grid size. Dispersion at larger scales is modeled explicitly by flow through a specific heterogeneous permeability field. We choose
high dispersivities of $4 d_{l} = 10 d_{t} = \Delta x = \Delta y$ with a factor 2.5 anisotropy.
Dispersion anisotropy, with transverse dispersivities generally lower than longitudinal ones, has the potential to further elongate viscous fingers. If transverse dispersion were high, it could stabilize the fingering \citep{tan1986stability}. Fickian diffusion is more pronounced in the transverse direction, which has the steepest compositional gradients. 

Unlike in most reservoir simulators, Fickian diffusion coefficients are not assumed to be a constant diagonal matrix, but are computed from irreversible thermodynamics as a full matrix of composition-dependent coefficients for each phase \citep{moortgatVII}. Example diffusion coefficients (and phase compositions) for a two-phase mixture of 1 mole of initial oil with 2 moles of injection gas are provided in Table~\ref{table::diffg}.

Figure~\ref{fig6} shows the overall CO$_{2}$ composition at 45\% PVI for simulations with and without gravity and with only Fickian diffusion or both diffusion and mechanical dispersion. The results are very close to those without dispersion in Figures~\ref{fig2} and \ref{fig4}, indicating that the flow is advection dominated. Whether flow is advection or dispersion dominated is often expressed in terms of a P\'eclet number, defined generically as
$
P_{e} = L u/D
$
with $u$ the advective velocity, $D$ a diffusion coefficient, and $L$ a characteristic length-scale. 

For fully compositional multiphase phase flow it is less straightforward to predict which flow mechanism dominates in terms of a P\'eclet number, because:
\begin{enumerate}
\item Advective phase velocities vary considerably throughout the domain due to changes in relative permeabilities (saturations) and viscosities (see Figure~\ref{fig3}). Flow could theoretically be advection dominated in one phase, and diffusion dominated in the other.
\item Fickian diffusion coefficients are composition dependent and fluxes depend on phase saturations and molar densities (Eq.~\ref{eq::difff}). Moreover the diffusive flux of each species is different with different coefficients (Table~\ref{table::diffg}) and different compositional gradients.
\item It is not clear what the characteristic length-scale is \citep{tan1986stability}. Some times the domain size in a particular direction is used, which may be appropriate for the convective flow, but for the dispersive component perhaps a finger-width or distance between fingers is more appropriate. For dispersion to be important in the context of fingering behavior, it does not have to transport species over the full domain width or height, but only from the fingers to the fluid in between to have a stabilizing effect.
\end{enumerate}
Nevertheless, by expressing a P\'eclet number in terms of either the dispersivities or the Fickian self-diffusion coefficient for CO$_{2}$ and taking either the domain height or width as the characteristic length it is clear that P\'eclet numbers are large. {\color{black}As an example, for the given injection rate of $7.3\%$ PV/yr, a stable displacement front (discussed below) propagates $~300\ \mathrm{m}$ in $\sim 14$ years, while viscous fingers already cover the full $600$ m width of the domain by that time, corresponding to characteristic advective velocities of ($1.4$--$2.8$) $\times 10^{-6}\ \mathrm{m}/\mathrm{s}$. The gas and oil diffusion coefficients in Table~\ref{table::diffg} are of order $10^{-10}$--$10^{-9}\ \mathrm{m}/\mathrm{s}^{2}$, so Fickian diffusion is important (P\'eclet number of $\le 1$) on scales of milli- to centimeters. Using either the domain width or height as the characteristic length-scale, P\'eclet numbers are $>10^{5}$. For mechanical dispersion, the longitudinal and transverse dispersivities are chosen as 10--25 cm, respectively. This implies P\'eclet numbers of $\sim 10^{3}$ (note that while the Fickian diffusion coefficients are computed self-consistently from the compositions, mechanical dispersion is essentially an up-scaling technique; dispersivities are user-defined and chosen to show the maximum effect of dispersion). }

\begin{figure}[!h]
\centerline{\includegraphics[width=.7\textwidth]{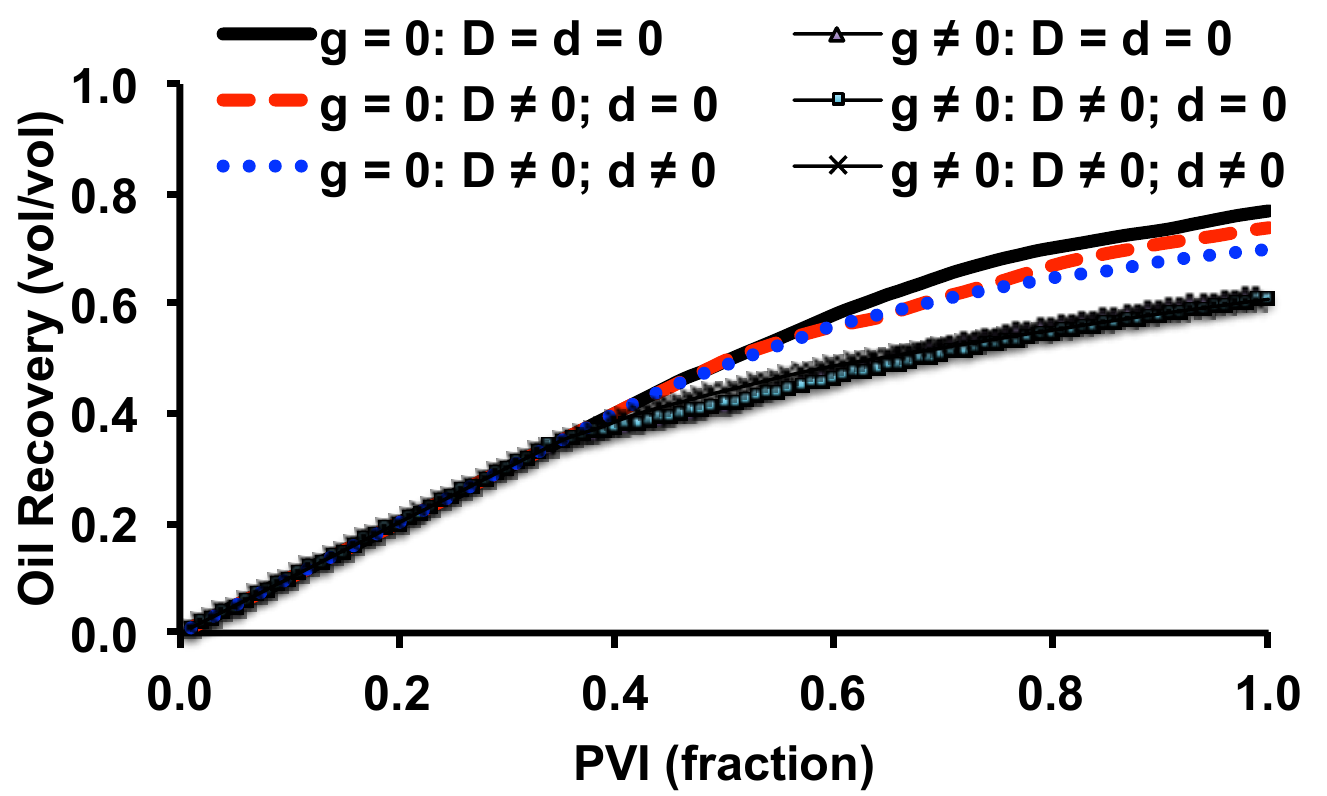}}
\caption{\label{fig7}Oil recovery for simulations in vertical ($g\ne0$) and horizontal ($g=0$) domains, with ($D=D^{\mathrm{Fick}} \ne 0$) and without ($D = 0$) Fickian diffusion and with ($d=d_{l}, d_{t}\ne 0$) and without ($d=0$) mechanical dispersion.}
\end{figure}

In the simulation with gravity, Fickian diffusion, and dispersion, the small-scale gravitational plumes are somewhat suppressed, but oil recoveries for all simulations with gravity are identical (Figure~\ref{fig7}). Oil recoveries for simulations without gravity but with Fickian diffusion and/or mechanical dispersion are slightly lower than without either, but this is most likely just due to random changes in fingering patterns.

More generally, diffusion can stabilize weak gravitational or viscous fingering, particularly at small scales (e.g., in laboratory experiments \citet{moortgatIV}). In carbon sequestration, for instance, gravitational fingering is triggered by a density difference in the aqueous phase of only about 1\%. Under those circumstances, stability analyses \citep{xu06,riaz,philipsequestration} and numerical simulations 
\citep{pruess3,pruesszhang, moortgatIII} show that gravitational fingers are triggered at a critical time ($t_{c}$) and with a critical wavelength ($\lambda_{c}$) given by
\begin{equation}\label{eq::tcrit}
t_{c} \propto D^{\mathrm{Fick}} \left(\frac{\mu \phi}{\mathrm{K} \Delta \rho g} \right)^{2}\quad\mbox{and}\quad
\lambda_{c} \propto D^{\mathrm{Fick}} \frac{\mu \phi}{\mathrm{K} \Delta \rho g}. 
\end{equation}
The critical (onset) time for instability can be very large for small $\Delta \rho$ and permeability. For viscous (and gravitational) fingering at the field scale, mobility ratios can easily be orders of magnitude (and $\Delta \rho$ of order one) and diffusion may rarely stabilize flow instabilities. \citet{zimmerman1992viscous,zimmerman1991nonlinear} also found that viscous fingering in single-phase flow is insensitive to dispersion for high P\'eclet numbers.
In the following simulations, diffusion and dispersion will be neglected, unless specified otherwise.

\subsection{Effect of Flow Rate}
It appears that the dependence of viscous fingering on injection rate has not been considered in the literature. To investigate whether the injection rate is important, simulation results are presented for rates of 1\%, 2.5\%, 5\%, 7.5\%, 10\% and 20\% PV/yr, and with and without gravity. 

\subsubsection{Without Gravity}
\begin{figure}[!h]
\centerline{\includegraphics[width=\textwidth]{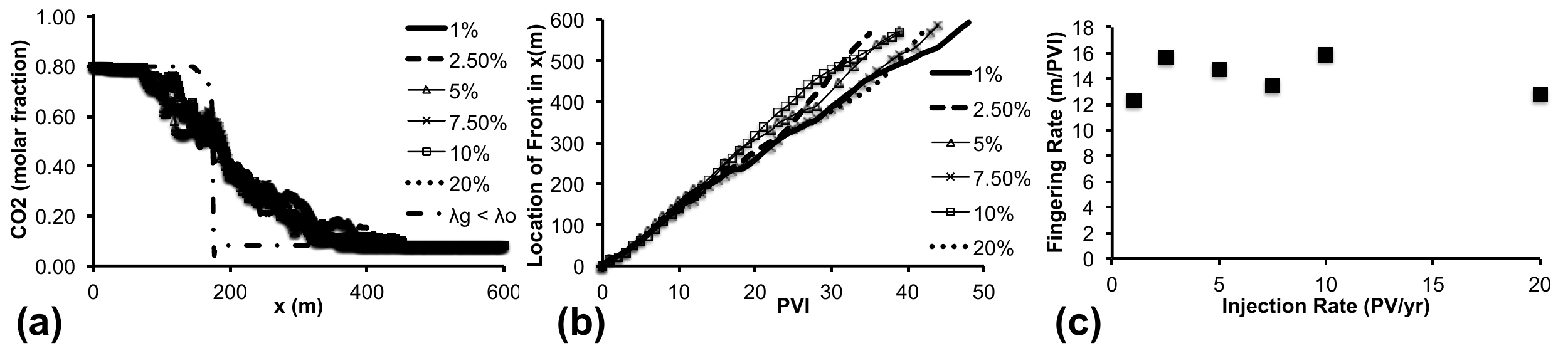}}
\caption{\label{fig8}Fingering growth in horizontal domain as a function of constant injection rates of $1\%$, $2.5\%$, $5\%$, $7.5\%$, $10\%$, and $20\%$ PV/yr.
Overall CO$_{2}$ molar fraction integrated over $y$ at $30\%$ PVI for all rates, as well as a front for stable displacement ($\lambda_{g}<\lambda_{o}$, dash-dotted) at 7.5\% PV/yr \textbf{(a)}, location of the fingering front versus PVI \textbf{(b)}; the slopes provide the PV/yr rate-normalized growth rates in m/PVI \textbf{(c)}. 
}
\end{figure}
The main interest is in the growth rate of \textit{viscous} fingering, so the first simulations are for a horizontal cross-section without gravitational effects.
To track the finger-tips the composition (or saturation) profiles are integrated over the $y$-direction. Darcy flow itself scales linearly with injection rate, so to investigate the degree of fingering relative to the average advective flow, results are plotted versus PVI, rather than time. {\color{black}Figure~\ref{fig8}a} shows averaged CO$_{2}$ profiles at $30\%$ PVI for all injection rates, as well as the front-location at 7.5\% PV/yr for \textit{stable} displacement (discussed in the next section). The CO$_{2}$ profiles have small local variations, but the averaged positions of the leading and trailing fronts are the same for all rates. 
{\color{black}Figure~\ref{fig8}b} shows the advance of the leading finger-tip versus PVI for all rates. Until $\sim 15\%$ PVI the curves are identical. At later times there is more variation, depending on whether one finger dominates or a few fingers grow to comparable sizes, but there is no trend in terms of injection rate. When the slopes of the middle panel are plotted versus injection rate it is clear that all growth rates fluctuate around 14 m/PVI. 

The conclusion is that the degree of viscous fingering does not appear to depend on injection rate. This is perhaps not surprising from Darcy's law: the pressure gradient is largely determined by the injection rate (for a given permeability), so regardless of mobility ratios, all (stable or unstable) phase velocities will scale with the injection rate. 

\subsubsection{With Gravity}
\begin{figure}[!h]
\centerline{\includegraphics[width=\textwidth]{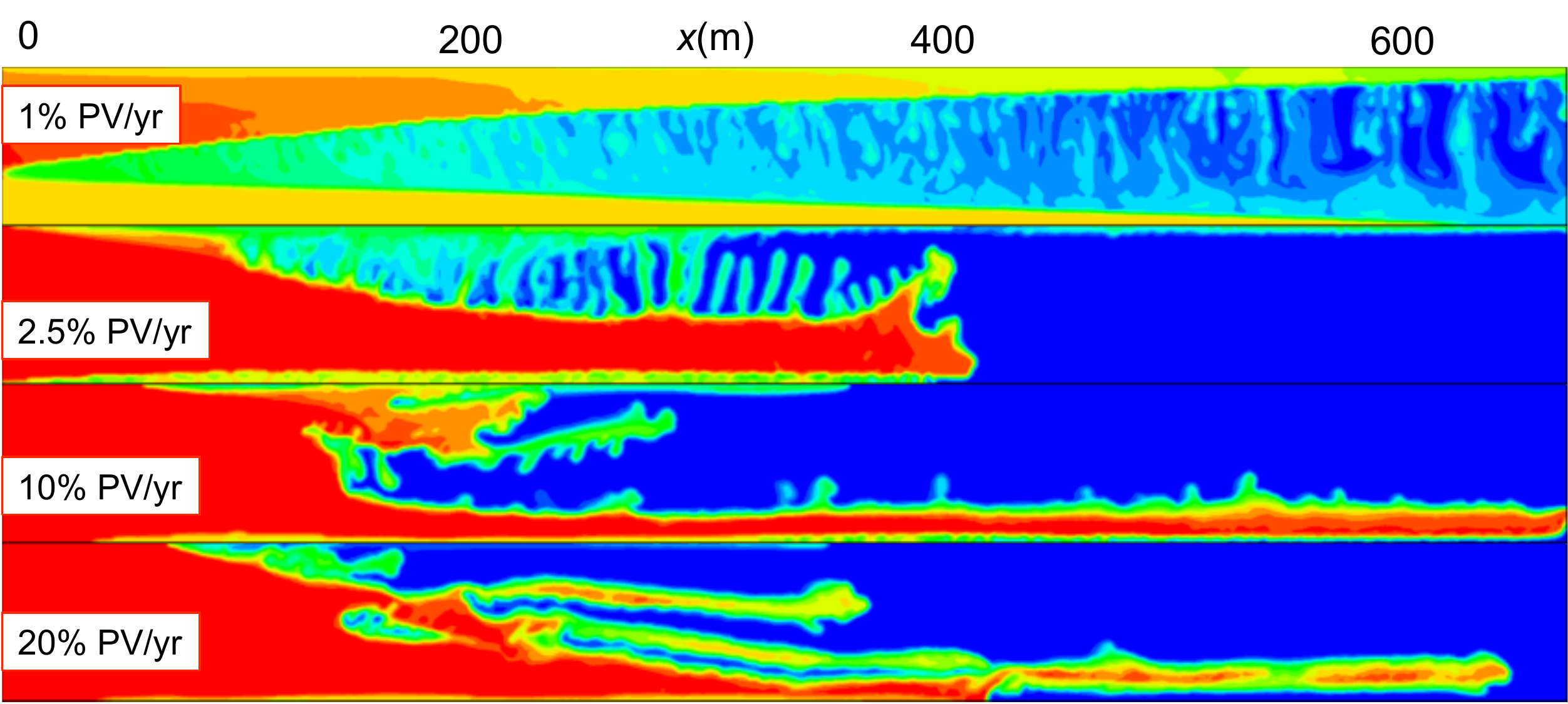}}
\caption{\label{fig8b}Overall CO$_{2}$ molar fraction in vertical domain at 35\% PVI for injection rates of $1\%$, $2.5\%$, $5\%$, $10\%$, and $20\%$ PV/yr (legend as in Fig.~\ref{fig4}).}
\end{figure}
The caveat to the above conclusion is, of course, that there were no processes competing with viscous flow. Fickian diffusion, capillarity, and gravity do not directly depend on injection rate (but mechanical dispersion does). At low injection rates, these processes may dominate, while at higher rates advective flow prevails. This is illustrated for gravitational flow in Figure~\ref{fig8b} at 35\% PVI. Given the large viscosity contrast and small density difference, gravity only clearly dominates for the lowest injection rate of 1\% PV/yr and viscosity becomes increasingly more important at the higher rates.

\subsection{Effect of Relative Permeability on Mobility Ratio}\label{sec::endpoints}
\begin{figure}[!h]
\centerline{\includegraphics[width=\textwidth]{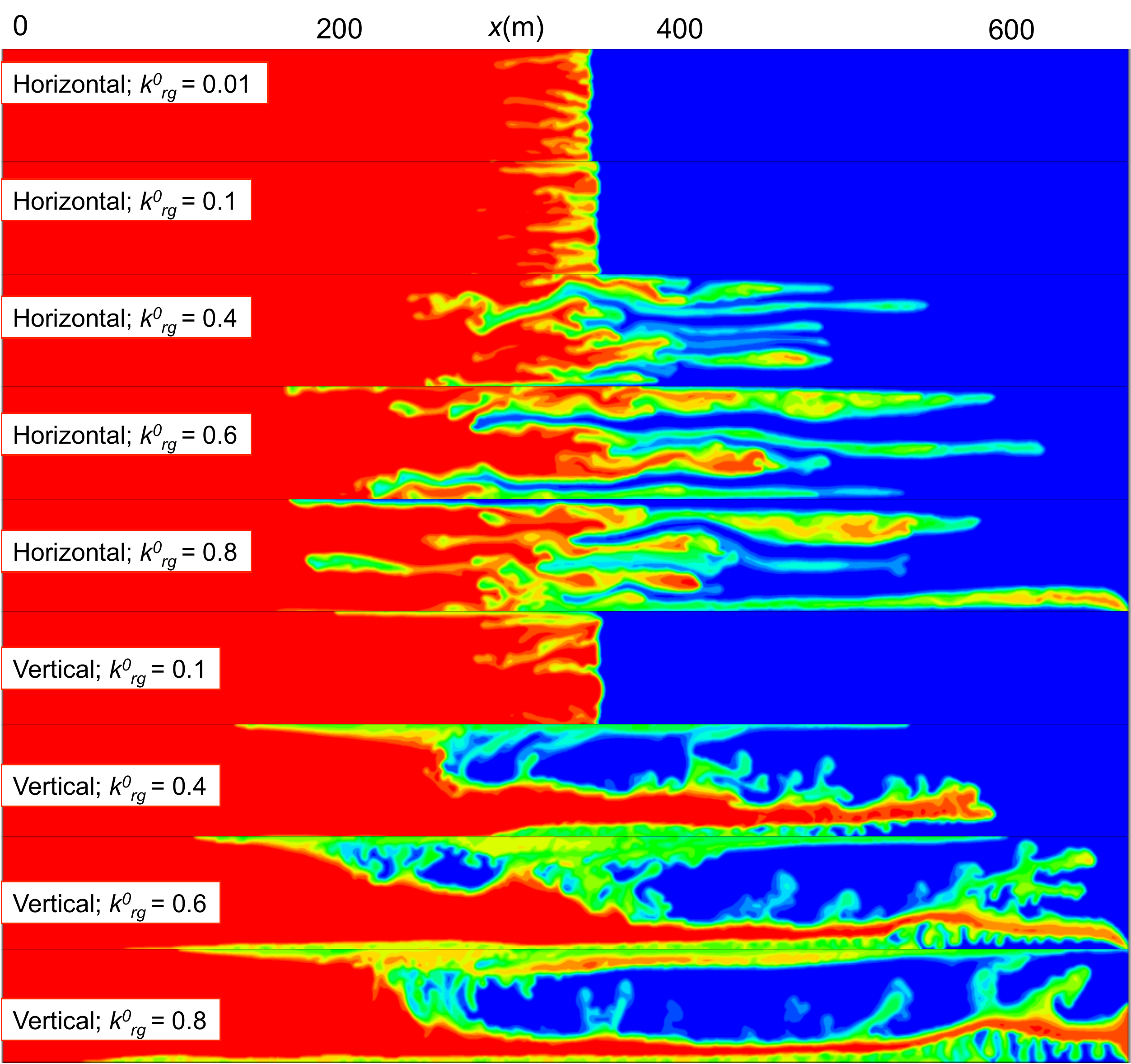}}
\caption{\label{fig9}Fingering growth at 50\% PVI, in vertical and horizontal domains (with and without gravity), as a function of mobility ratio, determined by the end-point relative permeabilities of gas ($k^{0}_{rg}=0.01$, 0.1, 0.4, 0.6, and 0.8).
}
\end{figure}
Next, the effect of mobility ratio on fingering growth is investigated. Phase viscosities are computed self-consistently from their compositions \citep{viscosity2}. It is undesirable to change viscosities artificially, because local viscosity changes due to phase behavior are one of the unique aspects of compositional multiphase flow that are the subject of this study. Instead, the {mobility} ratio is manipulated by lowering the end-point relative permeability for gas. The initial viscosity ratio is $\mu_{o}/\mu_{g}\sim 20$, so keeping the end-point relative permeability for oil at one, but setting that of gas to $0.05$ results in a mobility ratio (at end-point saturations) of unity. To study at what mobility ratios the flow becomes unstable to viscous fingering, simulations are performed for gas injection at 7.26\% PV/yr, with and without gravity, and for end-point relative permeabilities for gas of $k_{r,g}^{0}= 0.01$, $0.1$, $0.4$, $0.6$ and $0.08$. For completeness Fickian diffusion is also included, with typical diffusion coefficients as before (Table~\ref{table::diffg}).

Figure~\ref{fig9} shows CO$_{2}$ concentration profiles at 50\% PVI for different $k_{r,g}^{0}$, with and without gravity. It is clear that for $k_{r,g}^{0}\le 0.1$ the displacement front is essentially stable. This is reasonable, because behind the displacement front the gas viscosity has increased to $\sim 0.17$ cp due to compositional phase behavior, which makes the mobility ratio across the front $M^{\mathrm{front}} = \lambda_{g}/\lambda_{o} \approx 7.5 k_{r,g}^{0}<1$ for $k_{r,g}^{0}\le 0.1$. Note that without species transfer the viscosity ratio would remain at $21$ and stable displacement would require $k_{r,g}^{0}\le 0.05$.

Interestingly, for $M^{\mathrm{front}}<1$ gravitational effects are suppressed as well. The reason is that viscous fingers provided a larger interaction surface for vertical species exchange between oil and gas. The resulting local density changes triggered gravitational fingering. A higher gas mobility also allows gravitation fingers to propagate faster, which is apparent in Figure~\ref{fig9} for increasing $k_{r,g}^{0}$.

\begin{figure}[!h]
\centerline{\includegraphics[width=.85\textwidth]{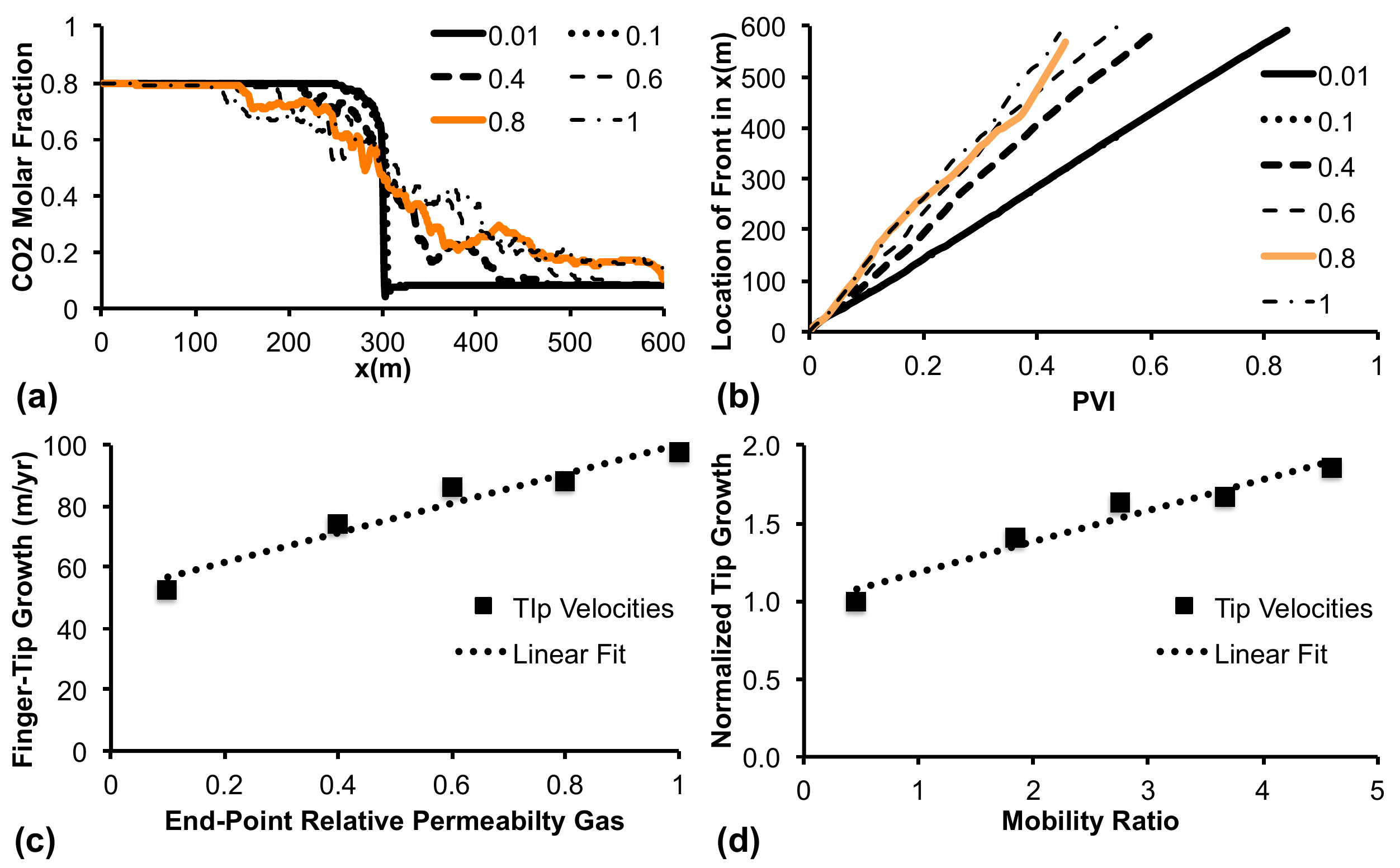}}
\caption{\label{fig9b}Finger growth w.r.t.~mobility ratio in a horizontal domain, determined by $k^{0}_{rg}=0.01$, 0.1, 0.4, 0.6, or 0.8. Integrated overall CO$_{2}$ composition profiles at 50\% PVI \textbf{(a)}, tip-location versus PVI \textbf{(b)}, tip-speed versus end-point relative permeability for gas \textbf{(c)} and normalized speed versus mobility ratio \textbf{(d)}.
}
\end{figure}

To further quantify the scaling of viscous fingering with mobility ratio, Figure~\ref{fig9b} shows the CO$_{2}$ concentration profiles integrated over the transverse direction, the location of the finger-tips as a function of PVI, and the associated tip propagation speed (slope of the former curves). At the given injection rate and domain size, piston-like stable displacement of all oil requires a propagation speed of the front at $43.6$ m/yr. For comparison, the front for $k_{r,g}^{0}=0.01$ - $0.1$ is $U^\mathrm{stable}= 52.8$ m/yr with breakthrough occurring at 82\% PVI.
These results suggest that the propagation speed of the fastest fingers ($U^\mathrm{tip}$) scales linearly with the mobility ratio (for a given injection rate and absolute permeability). Figure~\ref{fig9b} also shows the tip-speed, normalized by the stable displacement velocity, as a function of mobility ratio. The linear fit {\color{black} in Figure~\ref{fig9b}d} gives:
\begin{equation}\label{eq::tipspeed}
U^\mathrm{tip}/U^\mathrm{stable} \sim M^{\mathrm{front}}/5.
\end{equation}
Note that this should not necessarily be compared to fingering growth-rates obtained from linear stability analyses (e.g.,~\citet{tan1986stability} for single-phase flow). $U^\mathrm{tip}$ is the propagation speed of the largest fingers in the non-linear regime at late times after shielding and pairing has occurred. From mass conservation considerations it is likely that the tip-speed of the $n$ largest fingers, with an average width of $\Delta f$, scales with $L_{y}/(n\Delta f)$. However, it may not be straightforward to predict the late time finger-tip speeds from first principles, and Eq.~\ref{eq::tipspeed} is probably not general.

\subsection{Effect of Domain Aspect Ratio}
There has been some discussion in the literature on the effect of domain aspect ratio ($A = L_{x}/L_{y}$) on the development of viscous fingers (e.g., \citet{moissis1988simulation}). To investigate this effect for compositional multiphase flow, the domain width is reduced in the next set of simulations in $9$ increments of $60\ \mathrm{m}$ (in the $x$-direction), i.e., the domain sizes vary from the initial $600\ \mathrm{m}\times 60\ \mathrm{m}$ to $60\ \mathrm{m}\times 60\ \mathrm{m}$ ($A$ varying from 10 to 1). Injection is at a constant surface rate for all domains (which translates to $7.5\%$ PV/yr for the $600\ \mathrm{m}\times 60\ \mathrm{m}$ domain to $75\%$ PV/yr for the $60\ \mathrm{m}\times 60\ \mathrm{m}$ domain). 

\subsubsection{Without Gravity}
\begin{figure}[!h]
\centerline{\includegraphics[width=.7\textwidth]{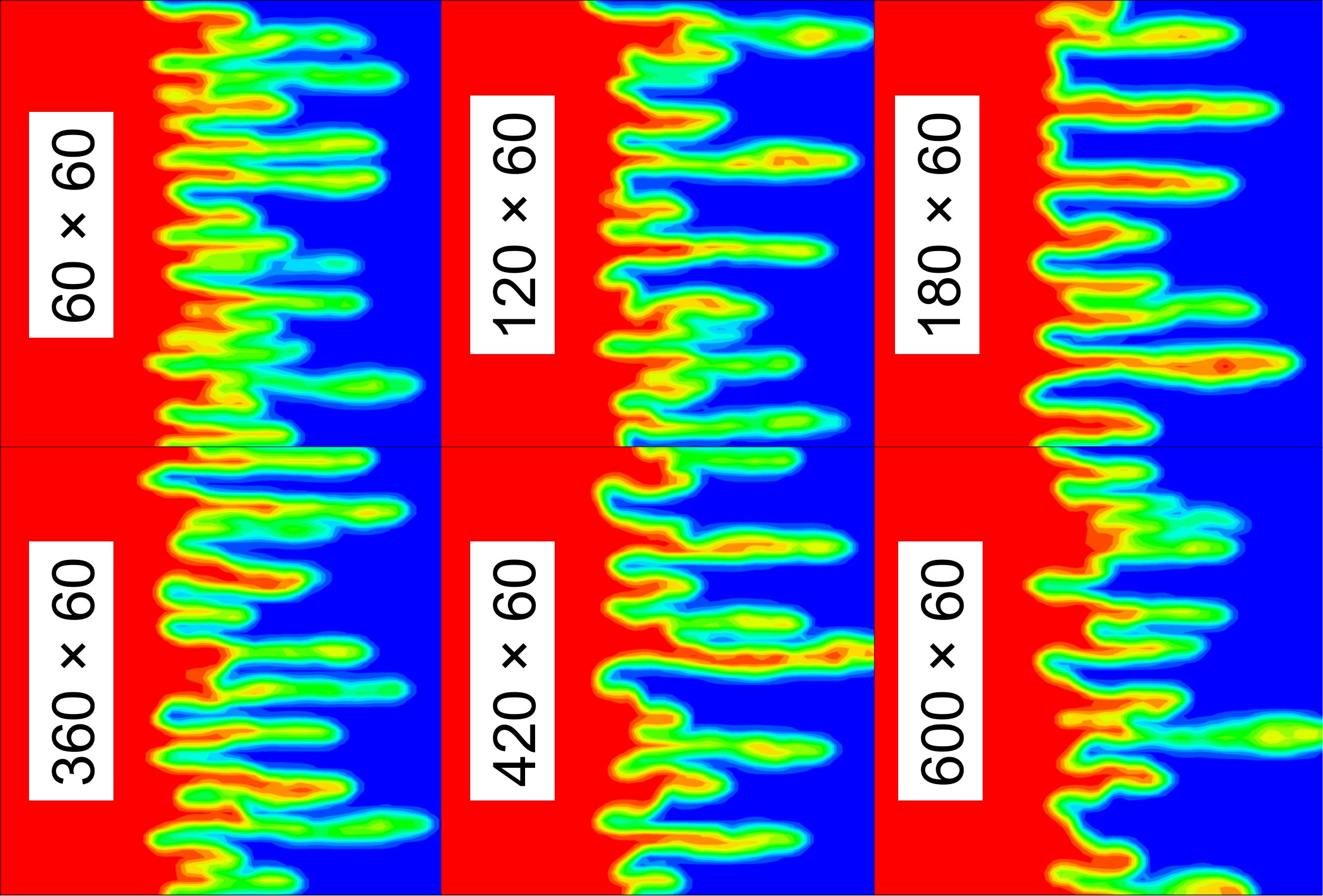}}
\caption{\label{figx}Overall CO$_{2}$ composition after 219 days of gas injection in the left-most $60\ \mathrm{m}\times 60\ \mathrm{m}$ of different (\textbf{horizontal}) domain widths of: $60$, 120, 180, 360, 420, and 600 m.}
\end{figure}
\begin{figure}[!h]
\centerline{\includegraphics[width=\textwidth]{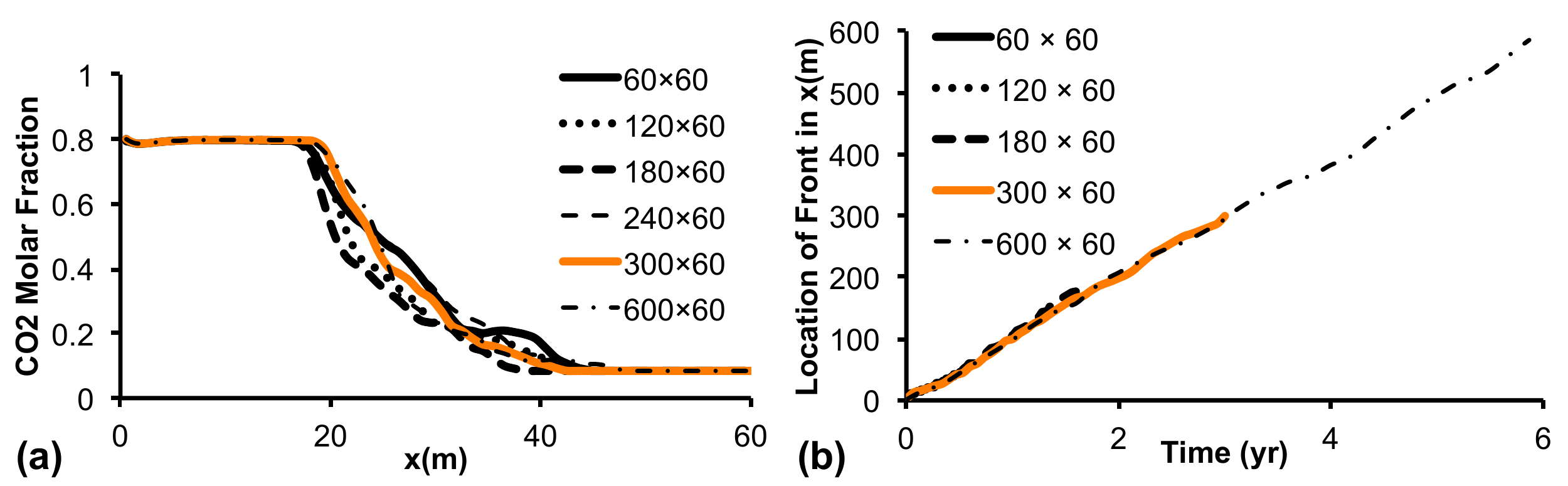}}
\caption{\label{figxx}Integrated CO$_{2}$ composition after 219 days in the left-most $60\ \mathrm{m}\times 60\ \mathrm{m}$ \textbf{(a)} and location of the front versus time \textbf{(b)} for different domain sizes. }
\end{figure}

To determine any dependency of the critical time and wavelength of the viscous instability on $A$, the early growth of fingering is compared in the left-most $60\ \mathrm{m}$ for each of the domain sizes after 219 days. From the six examples in Figure~\ref{figx} it appears that the onset time of fingering, number of fingers, and finger growth rate is independent of domain size until breakthrough. This is further demonstrated in Figure~\ref{figxx}, which compares integrated CO$_{2}$ compositional profiles after 195 days, as well as the average growth of the finger-tips throughout the full length of the domain, similar to Figure~\ref{fig8}. 
This suggests that the viscous fingering instability \textit{itself} is independent of $A$. It also means that the simulations are in the asymptotic regime of finger development \citep{zimmerman1992viscous}.

\subsubsection{With Gravity}
\begin{figure}[!h]
\centerline{\includegraphics[width=\textwidth]{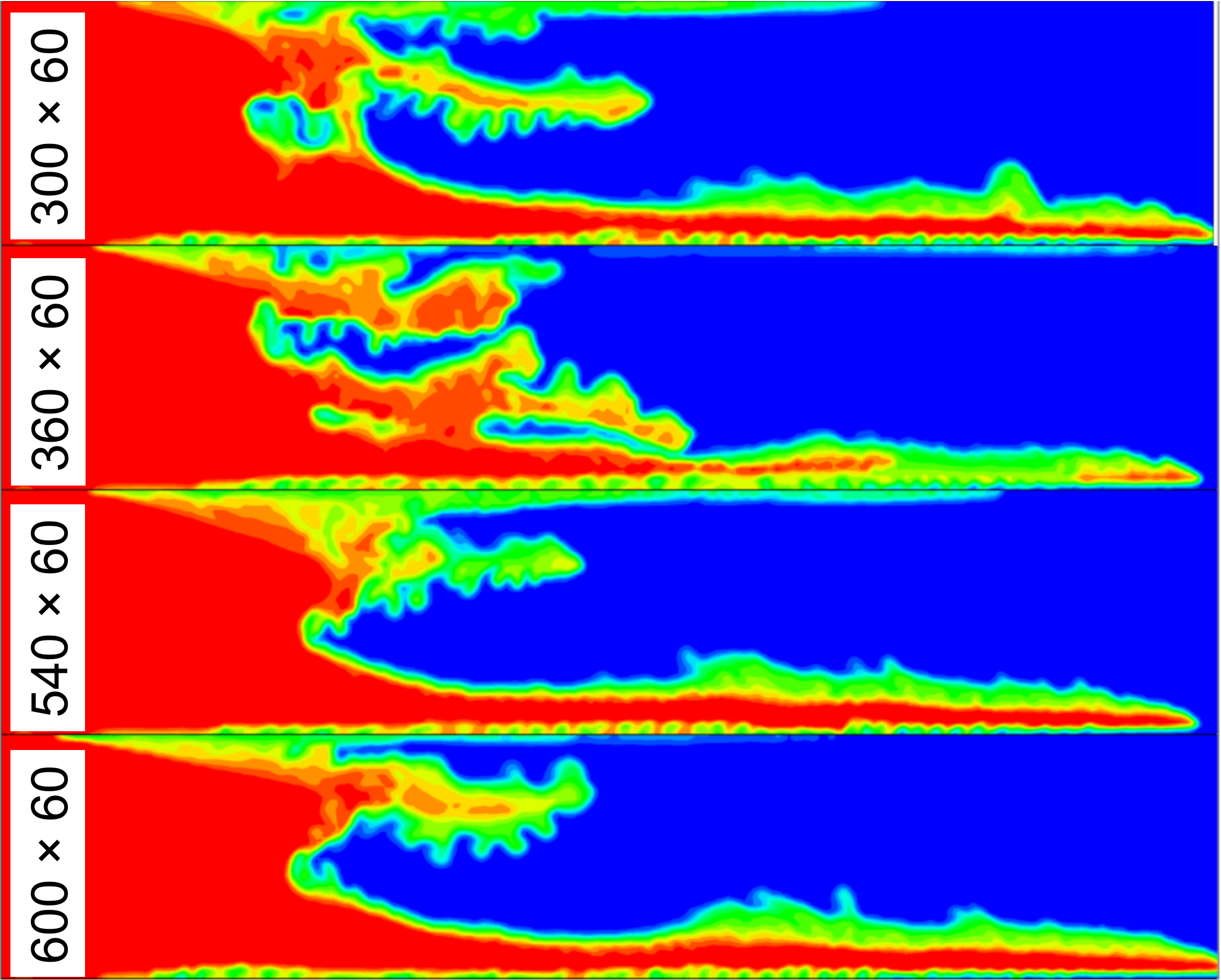}}
\caption{\label{fig11}Overall CO$_{2}$ composition after 4.8 years of gas injection in the left-most $300\ \mathrm{m}\times 60\ \mathrm{m}$ of different (\textbf{vertical}) domain widths of 300, 360, 540, and 600 m.}
\end{figure}
When gravitational effects are important, the domain aspect ratio has an effect in determining the relative importance of viscous flow in the horizontal direction and gravitational flow in the vertical direction. It appears, though, that even complex viscous plus gravitational flow is insensitive to $A$ \textit{before breakthrough}. Figure~\ref{fig11} shows the overall CO$_{2}$ composition after 4.8 years of gas injection in the left-most $300\ \mathrm{m}$ of four domains with different aspect ratios but a minimum of $L_{x}=300\ \mathrm{m}$ (to give gravitational patterns time to develop). The main flow patterns are remarkably similar given the highly non-linear flow (and different random permeability fields){\color{black}, demonstrating again that fingering behavior is insensitive to domain width.}

\subsection{WAG}
\begin{figure}[!h]
\centerline{\includegraphics[width=\textwidth]{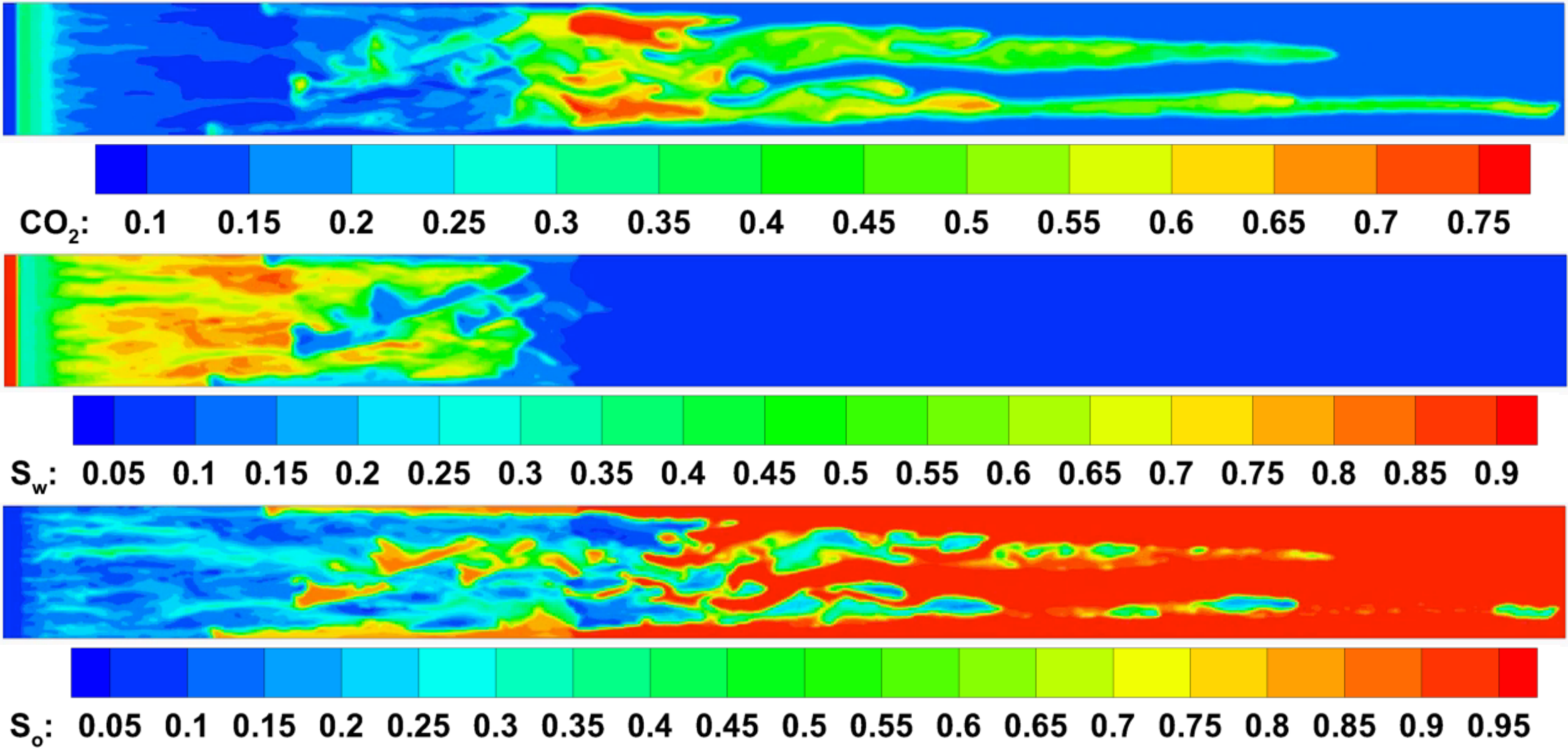}}
\caption{\label{wagnog}Overall CO$_{2}$ molar fraction (top), water (middle) and oil saturation (bottom) at 37\% PV of WAG injection in a \textbf{horizontal} domain.}
\end{figure}
WAG injection has been proposed as a strategy to improve mobility and sweep efficiency (e.g., \citet{christie19933d,juanes2006impact}). The basic idea is to inject slugs of water to reduce the gas mobility for a better sweep, alternated by slugs of gas to remove the residual oil left by water. {\color{black}We consider one WAG scenario} as an example of fingering in three-phase compositional flow when the slugs of gas are injected below the MMP. To the best of our knowledge, WAG injection has not been studied for three-phase compositional and compressible flow, except in our earlier work \citep{moortgatIII}. This example is an extension of that work. 

The set-up is the same as in the previous examples. Slugs of water (first) and gas (CO$_{2}$ and methane) are alternated at 6 month intervals with an injection rate of $7.3\%$ PV/yr. \citet{stone2} II three-phase relative permeabilities are assumed with $k_{r,o}^{0} = 0.5$ and $k_{r,w}^{0} = 0.3$, and exponents of 3. The residual oil saturation to water is 50\%. Gas-oil relative permeabilities are as before. Three-phase relative permeabilities change the process as compared to WAG with miscible gas slugs (two-phase).
The water viscosity is $0.48$ cp, so the water-oil viscosity ratio is 2.7, but the mobility ratio is 1.6.

\begin{figure}[!h]
\centerline{\includegraphics[width=\textwidth]{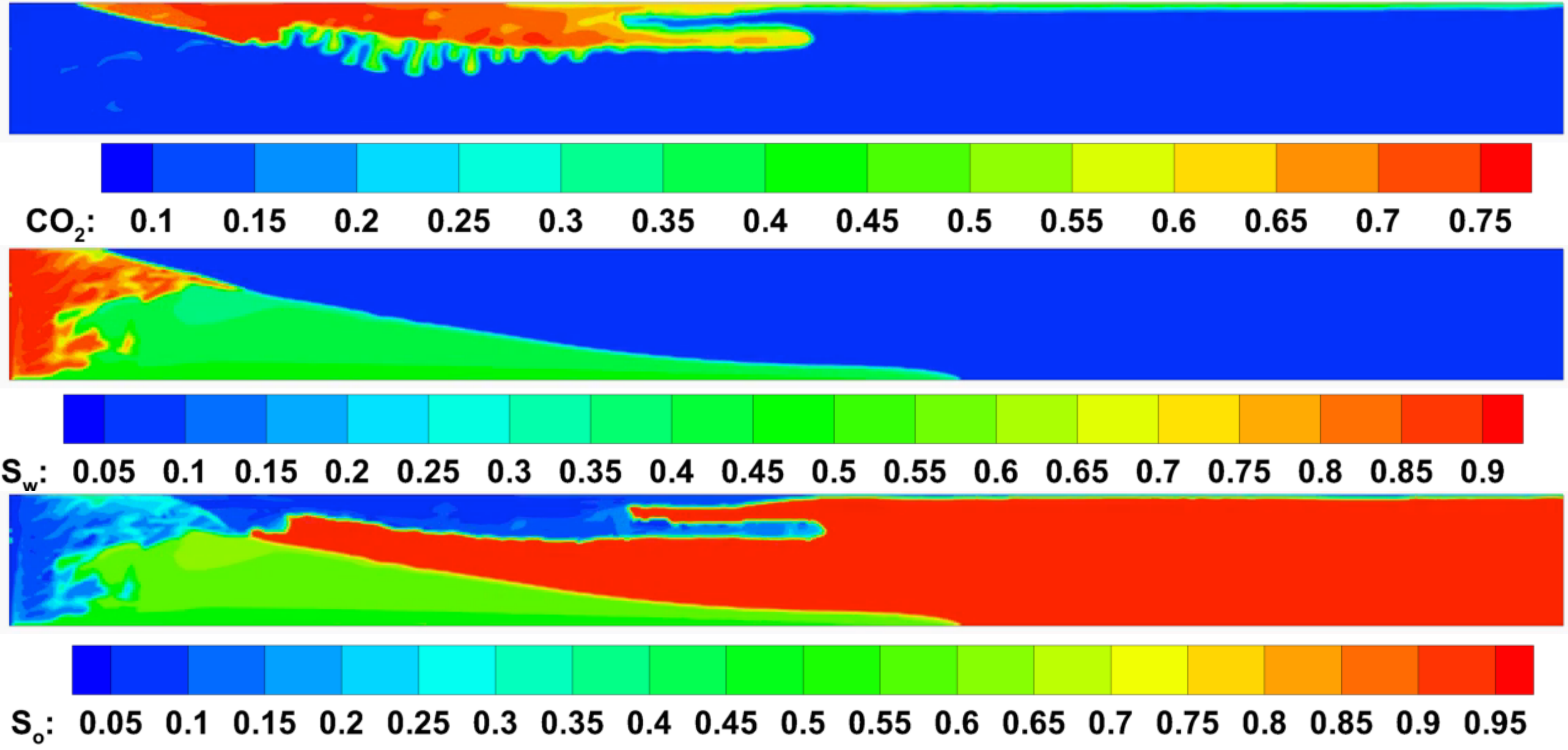}}
\caption{\label{wagg}Overall CO$_{2}$ molar fraction (top), water (middle) and oil saturation (bottom) at 37\% PV of WAG injection in a \textbf{vertical} domain.}
\end{figure}

\begin{figure}[!h]
\centerline{\includegraphics[width=.65\textwidth]{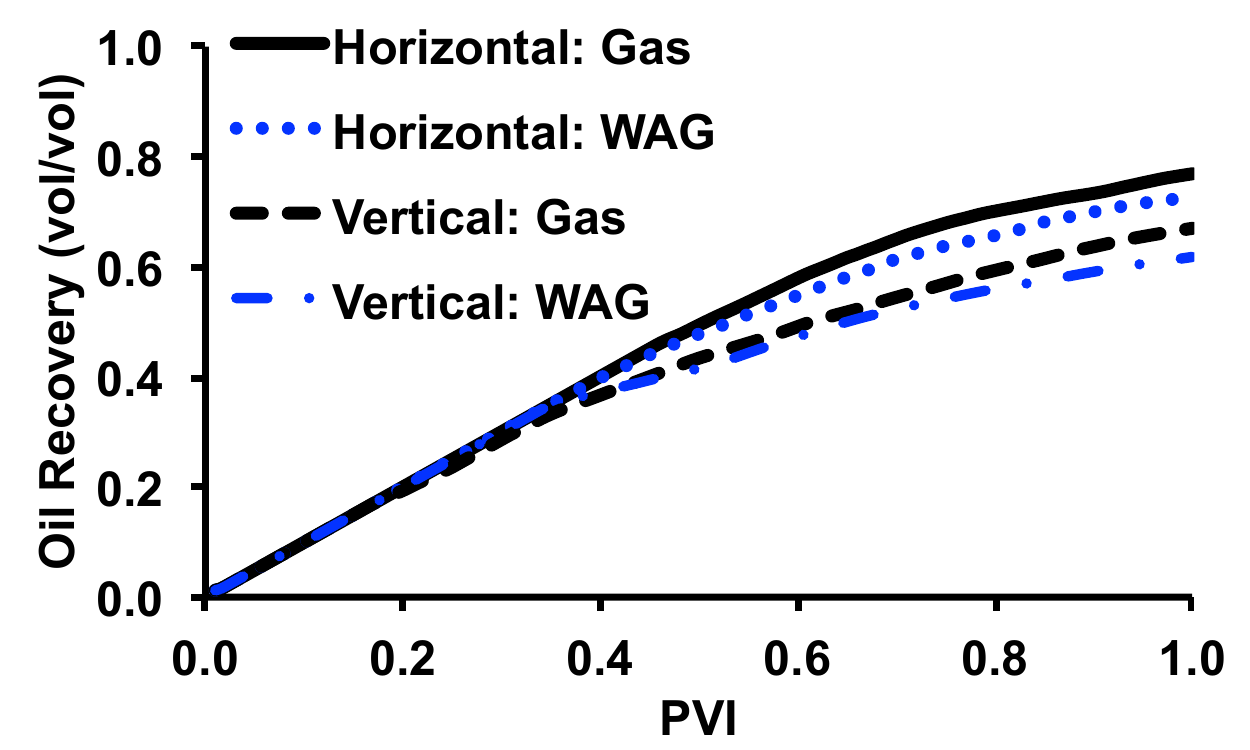}}
\caption{\label{wag}Oil recovery for gas and WAG injection in vertical and horizontal domains (with and without gravity).}
\end{figure}

Figures~\ref{wagnog} and \ref{wagg} show the overall CO$_{2}$ composition, water and oil saturations at the time of breakthrough at 37\% PVI, without and with gravity, respectively. Without gravity, the initial water slug provides a favorable stable displacement. However, the gas slugs have a higher mobility than both oil \textit{and} water, and viscous fingers of gas readily penetrate the water slugs. In fact, the breakthrough time is earlier for WAG (37\% PVI) than for only gas injection (this was also observed by \citet{christie19933d}). When gravity is included, water gravitationally segregates to the bottom with gravity override of gas in the top, which further deteriorates the sweep efficiency from WAG. Unless vertical permeability is low, this is of great concern.
Nevertheless, while WAG does not {improve} the sweep efficiency, the oil recoveries (Figure~\ref{wag}) are surprisingly similar (both without and with gravity) at only half the gas requirement. As such, WAG still provides a significant economic advantage. 

\subsection{Effect of Correlated Heterogeneity}
The effect of correlated permeability fields on fingering in miscible flow has been investigated by several authors (e.g., 
\citet{tan1992viscous,moissis1993simulation,tchelepi1994interaction,chang1994co2}). The degree to which those conclusions hold true for fingering in compositional multiphase flow is another subject that warrants a careful analysis. As a preliminary investigation, the next set of
8 simulations considers correlated permeability distributions, generated with the open source multivariable geostatistics package \texttt{gstat} \citep{gstat}. The {\color{black}(scalar)} permeabilities are given a correlation length equal to the domain height (60 m), an anisotropy of 50\% lower permeability in the vertical direction, and a variance of $0.25$, $0.5$, $0.9$, or $1.8$ times the average permeability. Simulations are done for each permeability field with and without gravity and including both Fickian diffusion and mechanical dispersion with the same coefficients as in 
Section~\ref{sec::diff}. Diffusion is generally more pronounced in heterogeneous (particularly, layered or fractured) reservoirs, because compositional gradients tend to form between regions of different permeabilities. 

\begin{figure}[!h]
\centerline{\includegraphics[width=\textwidth]{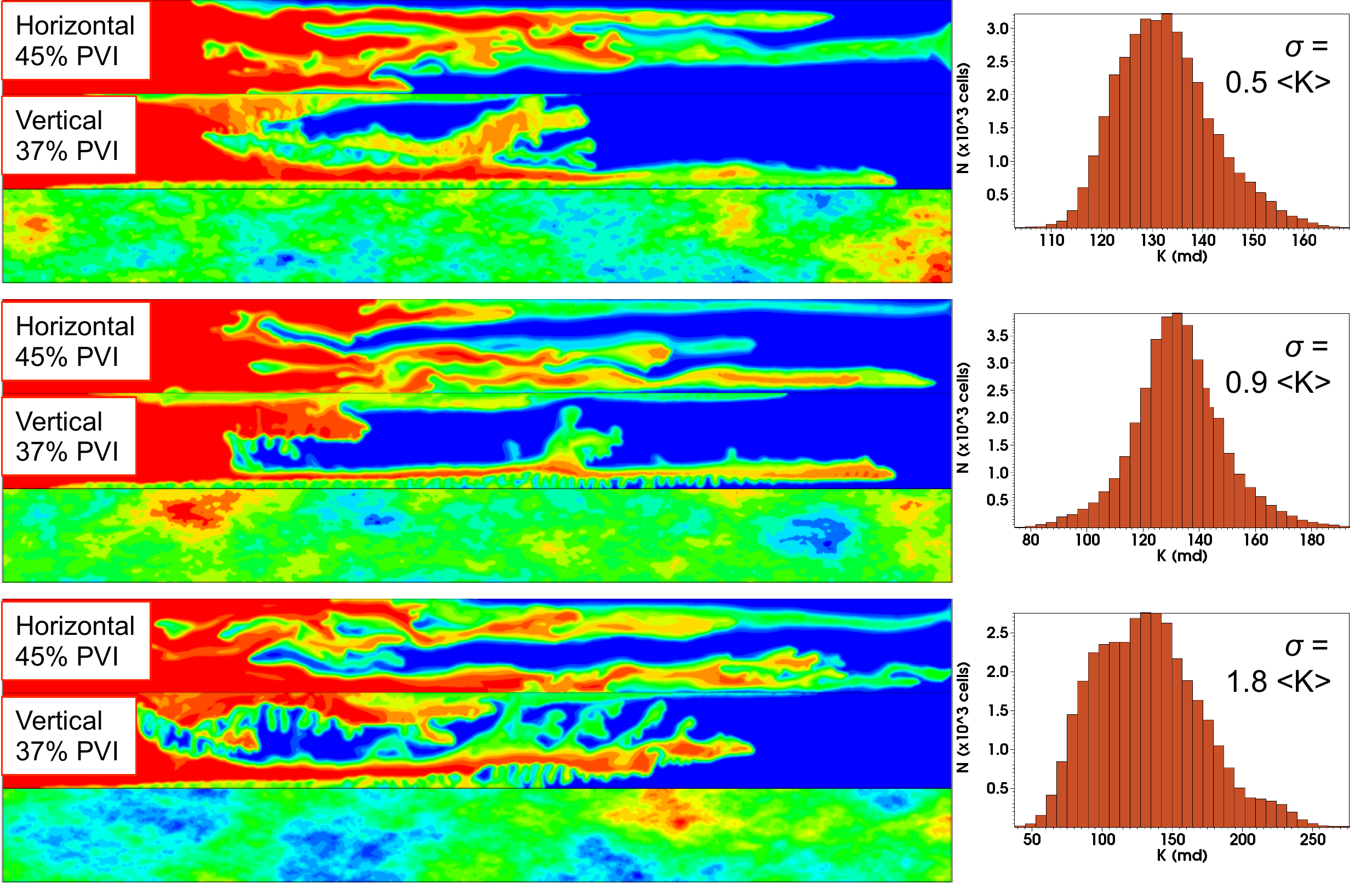}}
\caption{\label{correlated}Overall CO$_{2}$ molar fractions for simulations in vertical (at 37\% PVI) horizontal (at 45\% PVI) domains with three different correlated permeability fields. The permeability distributions with variances of $0.5$, $0.9$, and $1.8\times <\mathrm{K}>$ are shown both as color plots and histograms.}
\end{figure}

For the lowest variance the results are nearly indistinguishable from the previous uncorrelated permeability distribution. Figure~\ref{correlated} shows results for the three remaining geostatistical models. For the horizontal domains the viscous fingering is not affected significantly, even for the permeability distribution with the highest variance, which ranges from $38$ to $277$ md. Breakthrough still occurs around $45\%$ PVI. For the vertical cross-sections, gravitational flow is suppressed and breakthrough delayed, but this is mostly because the effective vertical permeability is reduced by a factor of two.

In \textit{layered} sedimentary formations with low effective vertical permeabilities \textit{gravitational} fingering tends to be stabilized, but \textit{viscous} fingering within individual (near-) horizontal layers with relatively uniform permeabilities may be similar to (2D) results presented in this work. 

\subsection{Fingering in Three-Dimensional Flow}
This final example investigates the degree to which viscous and gravitational fingering may change in three-dimensional flow. {\color{black}All other parameters (such as injection rates in PV/yr) are the same as in the 2D examples.}
Results are presented for a $600\ \mathrm{m}\times 12\ \mathrm{m}\times 60\ \mathrm{m}$ domain, such that about three $4$ m-wide fingers may form in the third dimension. The grid has the same resolution as before in the plain orthogonal to the main fingering flow ($\Delta y = \Delta z = 1$ m). To reduce computational cost, the $x$-dimension is discretized by 300 elements, but the grid is refined linearly in $x$ such that $\Delta x \sim 1$ m in the first $\sim 100$ m of the domain to resolve the small-scale onset of fingering, with larger grid cells near the production well where the fingers have grown longer. This grid has $216,000$ elements, which is quite computationally costly for $9$-component compositional two-phase flow. Note that the higher-order FE methods can provide high accuracy on relatively coarse grids, while lowest-order finite volume simulations would require considerably finer grids to obtain comparable results.

\begin{figure}[!h]
\centerline{\includegraphics[width=\textwidth]{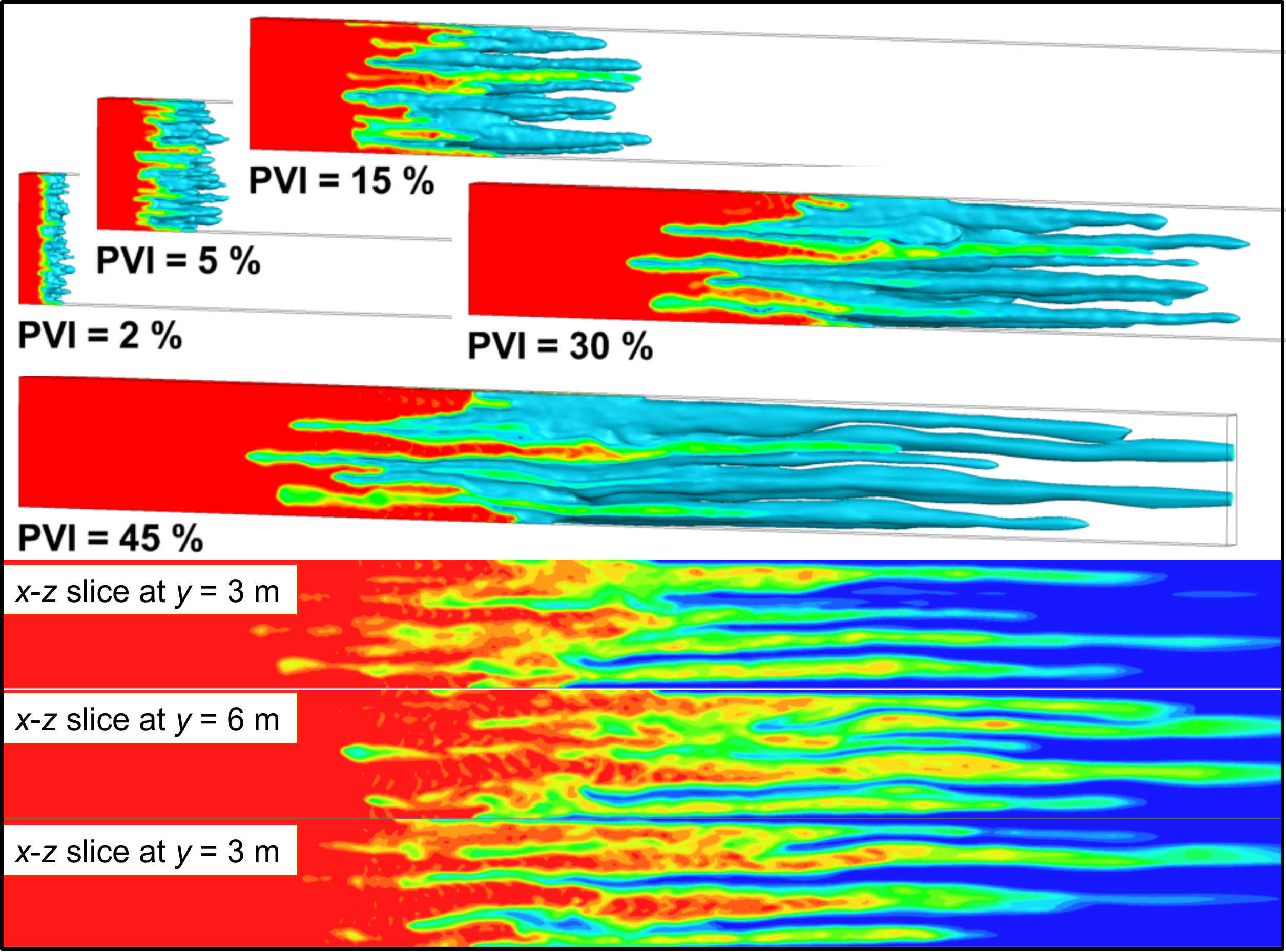}}
\caption{\label{fig19}Overall CO$_{2}$ molar fraction for simulations on $600\ \mathrm{m}\times12\ \mathrm{m}\times 60\ \mathrm{m}$ domain with Fickian diffusion and mechanical dispersion, but \textbf{without gravity}. 20 mol\% CO$_{2}$ iso-surfaces are shown for 2\%, 5\%, 15\%, 30\% and 45\% PVI.
Bottom panels show CO$_{2}$ molar fraction at 45\% PVI on 2D $x$-$z$-cross-sections for $y=$ 3, 6, and 9 m. Color-scale is as in Figure~\ref{fig2}.}
\end{figure}

First gravity is neglected, which for a 3D domain can be interpreted as a low vertical permeability. Figure~\ref{fig19} shows iso-surfaces of CO$_{2}$ composition at different times in 3D as well as on three $x$-$z$ cross-sections at 45\% PVI. The results are remarkably similar to those for 2D simulations, with the same breakthrough time and oil recovery (Figure~\ref{fig21}). This is likely because flow is predominantly in the $x$-direction (lower dimensional). The implication is that 2D simulation results may also be generalized to 3D flow.

\begin{figure}[!h]
\centerline{\includegraphics[width=\textwidth]{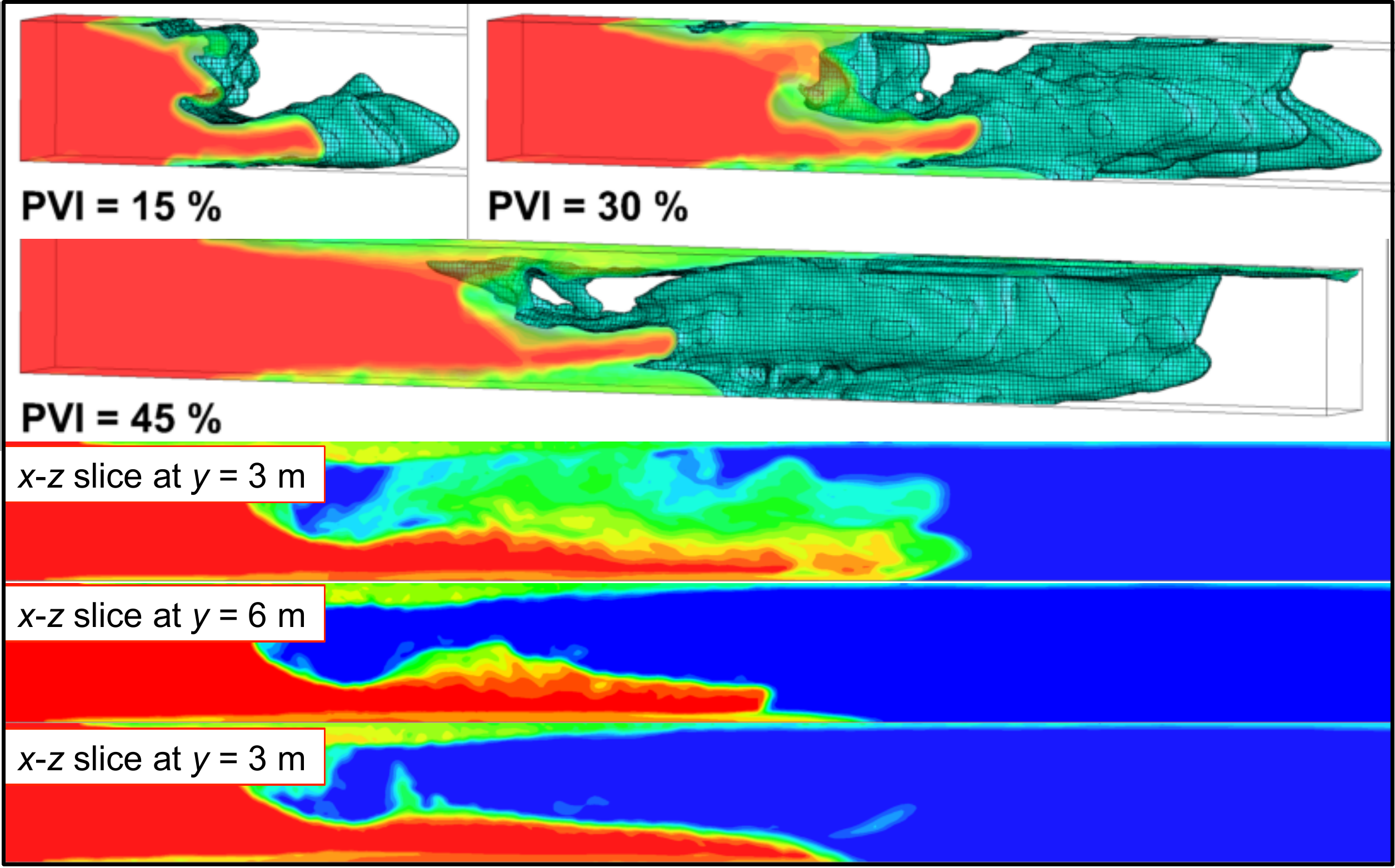}}
\caption{\label{fig20}Overall CO$_{2}$ molar fraction for simulations on $600\ \mathrm{m}\times12\ \mathrm{m}\times 60\ \mathrm{m}$ domain with Fickian diffusion, mechanical dispersion, and \textbf{with gravity}. 20 mol\% CO$_{2}$ iso-surfaces are shown for 15\%, 30\% and 45\% PVI.
Bottom panels show CO$_{2}$ molar fraction at 30\% PVI on 2D $x$-$z$-cross-sections for $y=$ 3, 6, and 9 m. Color-scale is as in Figure~\ref{fig2}.}
\end{figure}

When gravity is included, the results are less straightforward (Figure~\ref{fig20}): breakthrough is slightly later in 3D than in 2D. This is the opposite of what was found by \citet{tchelepi1994interaction}. In that work, though, a much higher density contrast was considered and 3D flow was more sensitive to gravity override. In this example, the density difference is insufficient to cause gravity override and the main effect of gravity is to cause perturbations of the viscous flow in the transverse (vertical) direction. In all 2D simulations (Figures~\ref{fig4}, \ref{fig6}, \ref{fig8b}, and \ref{fig9}) this vertical flow component caused the merging and destruction of viscous fingers at intermediate $z$, but allowed two dominant fingers to remain in the top and bottom of the domain. Those fingers resulted in early breakthrough and low oil recovery. 

In 3D, there are more flow paths available and it appears that this allows more mixing in the vertical direction, which delays the breakthrough of the top and bottom fingers. Nevertheless, those two dominant fingers are still present in 3D. Several of the $x$-$z$ cross-sections show similar flow patterns as in 2D. The final oil recoveries are also the same. This suggests that 2D simulations for viscosity dominated flow, with small density effects, may be (approximately) applicable to 3D with gravity as well. Additional 3D simulations were performed for a $600\ \mathrm{m}\times 60\ \mathrm{m}\times 60\ \mathrm{m}$ domain on a $300\times30\times 30$ grid and resulted in nearly identical oil recoveries.

\begin{figure}[!h]
\centerline{\includegraphics[width=.7\textwidth]{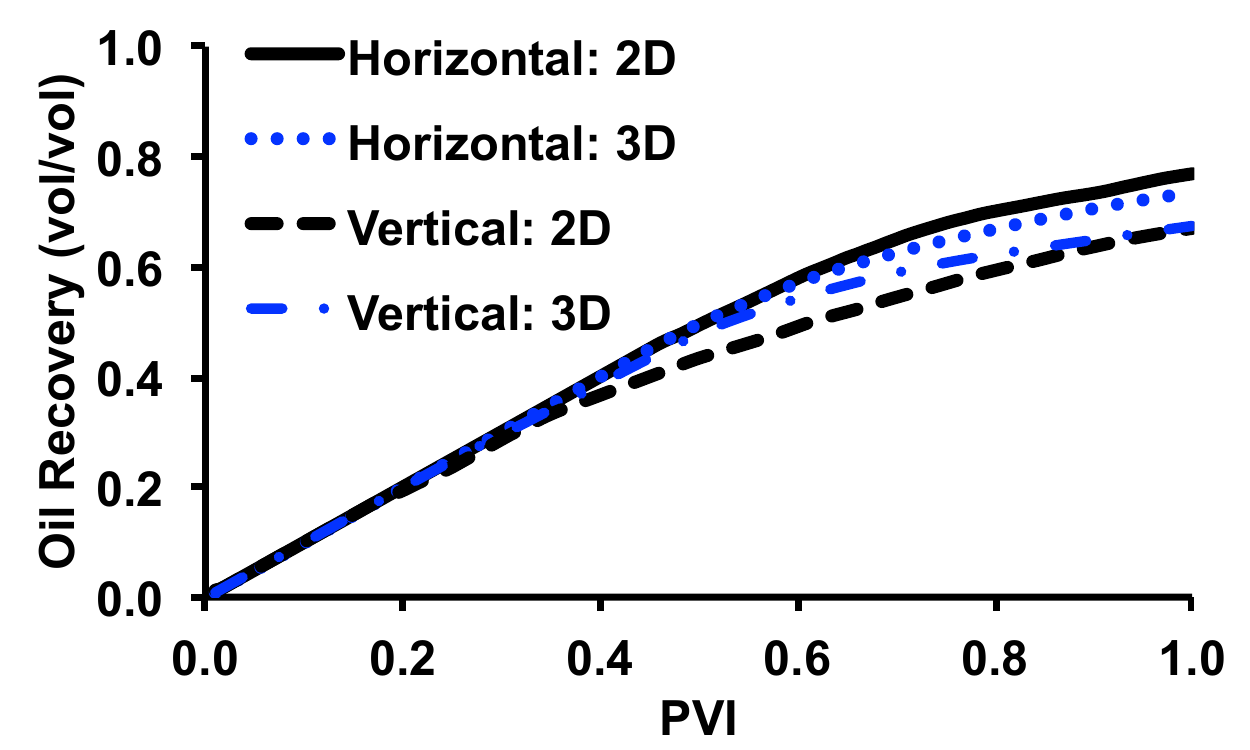}}
\caption{\label{fig21}Oil recovery for fingering in 2D and 3D simulations with and without gravity.}
\end{figure}

\section{Conclusions}
This work takes a first detailed look at interacting viscous and gravitational fingering in compositional and compressible multiphase flow with considerable mass transfer between the phases and associated phase behavior. The study is for high P\'eclet numbers, and initial gas-oil viscosity ratio and density difference of $21$ and $0.7\%$, respectively.
The similarities and differences with respect to earlier studies that considered miscible single-phase, or immiscible two-phase, flow are summarized here. 
\begin{enumerate}
\item[] \textbf{Mechanical Dispersion} A model for anisotropic velocity dependent mechanical dispersion was implemented for this work. As in previous studies \citep{zimmerman1992viscous,zimmerman1991nonlinear} it is evident that for typical anisotropic dispersivities, mechanical dispersion does not affect fingering behavior if the P\'eclet number is high (Figure~\ref{fig6}). 
\item[] \textbf{Fickian Diffusion} Diffusion in compositional multiphase flow is modeled with a full matrix of composition dependent coefficients in each phase (e.g., Table~\ref{table::diffg}). Fickian diffusion is driven by chemical potential gradients between gas in fingers and the surrounding oil and acts as a restoring force for viscous and gravitational fingering. Diffusion can stabilize such instabilities at small scales (e.g., \citet{moortgatIV, shahraeeni2015high}) or for weak instabilities, such as gravitational fingering in carbon sequestration from a 1\% density contrast (Eq.~\ref{eq::tcrit}). However, for strong viscous instabilities with high P\'eclet numbers at large scales, the diffusive time-scale is too long to significantly reduce the degree of fingering (Figure~\ref{fig6}).
\item[] \textbf{Mobilities} For compositional multiphase flow, the initial viscosity ratio between fluid-in-place and displacing fluid does not necessarily predict whether the flow will be unstable, for two reasons: 1) phase viscosities may change significantly due to species exchange and associated phase behavior, and 2) mobilities in two-phase flow also depend on relative permeabilities, which may vary considerably depending on wettability and saturation history (Figures~\ref{fig9}, and \ref{fig9b}). This study suggests that the tip-speed of the leading fingers scales linearly with the mobility ratio.
\item[] \textbf{Flow Rate} The dependence of fingering on injection rate is investigated for the first time. Without gravitational effects, the tip-speed of the leading fingers scales linearly with the average flow rate (Figure~\ref{fig8}). This suggests that the \textit{degree} of viscous fingering is not sensitive to injection rate, and average gas concentration profiles at a given PVI are nearly identical.
When gravity competes with viscous flow, lower injection rates result in more profound gravitational effects (Figure~\ref{fig8b}). 
\item[] \textbf{Gravity} This work focuses on viscous fingering by considering a displacing fluid with nearly identical density to the displaced fluid. In compositional multiphase flow, though, species transfer between the phases can change local densities to $>10\%$, which alters the fingering behavior significantly. Such effects would be absent in earlier work on fingering in incompressible or immiscible flow. 
\item[] \textbf{Domain Size and Aspect Ratio} Domain sizes are chosen such that the viscous fingers are in the asymptotic regime (infinite horizontal extent). The degree of fingering does not appear to depend on domain aspect ratio before breakthrough.
\item[] \textbf{WAG} Water-alternating-gas injection has been proposed to improve mobility and sweep efficiency, but has not be studied in detail when the gas slugs are injected below the MMP. An example with a unit WAG ratio shows that for typical three-phase relative permeabilities, gas fingers readily penetrate the water slugs. Additionally, when the vertical permeability is high, water tends to gravitationally segregate to the bottom with gravity override of gas in the top. For the simulated conditions, WAG does not reduce the degree of fingering nor improve the sweep efficiency, but it \textit{does} result in oil recoveries similar to only gas injection at only half the (costly) gas requirement.
\item[] \textbf{Correlated Heterogeneity} At high P\'eclet numbers, viscous fingering does not appear to be sensitive to permeability distributions with a long correlation length and a variance of up to $\sim 180\%$. Anisotropy with a reduced vertical permeability reduces gravitational effects as expected.
\item[] \textbf{Dimensionality} Three-dimensional simulations of viscous fingering with only weak gravitational effects are found to be remarkably similar to two-dimensional results because the flow is predominantly in the horizontal ($x$-)direction. Most 2D results presented in this work may therefore apply to 3D as well. In earlier work, we found that for gravitational fingering with flow predominantly in the vertical direction, 2D and 3D simulations also provide reasonably similar results \citep{shahraeeni2015high}.
When density contrasts are larger and viscous (horizontal) and gravitational (vertical) fluxes are of comparable magnitudes, three-dimensional simulations may be unavoidable \citep{tchelepi1994interaction}. 

\end{enumerate}

  \begin{table*}[htdp]
 \caption{\label{table::fluid}Oil and injection gas mass densities and viscosities (from \citet{viscosity2}) at initial compositions, and for a two-phase mixture of 1 mole initial oil with 3 moles gas.}
 \begin{center}
 \begin{tabular*}{\hsize}{@{\extracolsep{\fill}}l|cccc}\hline
& $\rho_{o}$ & $\rho_{g}$ & $\mu_{o}$ & $\mu_{g}$ \\\hline
Initial: & $736\ \mathrm{kg}/\mathrm{m}^{2}$	&  $731\ \mathrm{kg}/\mathrm{m}^{2}$	&   1.28 cp & 0.06 cp  \\
Mixture: & $818\ \mathrm{kg}/\mathrm{m}^{2}$	&  $647\ \mathrm{kg}/\mathrm{m}^{2}$	& 1.09 cp   &  0.09 cp \\
 \hline
 \end{tabular*}
 \end{center}
 \end{table*}

 \begin{table*}[htdp]
 \caption{\label{table::diffg}Phase compositions ($x_{g,i}$ and $x_{o,i}$ in mol\%) and effective diffusion coefficients ($\phi D_{g,ij}\times 10^{-10}\ \mathrm{m}^{2}/\mathrm{s}$ and $\phi D_{o,ij}$ $\times 10^{-10}\ \mathrm{m}^{2}/\mathrm{s}$) for 1 mole of initial oil mixed with 2 moles of injection gas, {\color{black}with $\phi$ the porosity. A further reduction due to tortuosity is not considered.}}
 \begin{center}
 \begin{tabular*}{\hsize}{@{\extracolsep{\fill}}ll|ccccccccc}\hline 
$x_{g,i}$ && CO$_{2}$ & C$_{1}$ & C$_{2}$ & C$_{3}$ & C$_{4-5}$ & C$_{6-9}$ & C$_{10-14}$ & C$_{15-19}$  \\\hline
56\%: &CO$_{2}$ &  12.44   &    -7.46   &    -5.78   &    -4.22    &  -2.86   &    -1.21    &   0.59    &    4.83 \\
32\%: &C$_{1}$   & -26.19	&    -6.08  &     -27.70 &    -26.06 &       -24.54 &       -21.68  &    -19.47   &    -21.13\\
2.2\%:&C$_{2}$ &  2.30  &      2.15    &    23.57   &     2.09   &     2.05    &    1.90   &    1.68    &    1.42 \\
1.3\%:&C$_{3}$ &  1.99   &     1.92    &    1.84   &    20.68   &     1.75  &      1.59    &    1.39  &     1.24 \\
1.1\%:&C$_{4-5}$ &   2.24 &        2.20 &        2.09 &       2.04 &        18.78 &        1.77    &    1.54   &     1.39 \\
1.9\%: &C$_{6-9}$ & 5.71     &  5.70    &    5.35    &    5.20    &    4.00   &     19.04    &    3.88   &     3.52\\
1.3\%: &C$_{10-14}$ &  5.72   &     5.73   &     5.36   &     5.21    &    5.01 &        4.49   &     17.15    &    3.48\\
0.7\%: &C$_{15-19}$ &  3.83 &        3.78     &  3.57     &   3.48 &        3.37 &        3.06   &     2.66   &     15.12\\\hline
$x_{o,i}$ && &&&&&&&\\\hline
52\%: &CO$_{2}$ &  	 5.07    	&     -6.11   	&   	-4.84 	&    	-4.02  	&  	-3.29   	&    	-2.27    	&   	-1.40    	&    	-0.55 		\\
26\%: &C$_{1}$   &  	-4.64		&      7.11  	&     	-4.85 	&    	-4.42 	&      -4.03 	&      -3.46  	&    	-3.12   	&    	-3.67	\\
2.4\%: &C$_{2}$ &     	3.46    	&     -4.79  	&    	11.29  	&     	 0.02		&    	0.04    	&    	0.06  	&    	0.07    	&    	0.07		\\
1.6\%: &C$_{3}$ &     	6.94    	&      1.61  	&    	 0.03		&    	10.04  	&     	0.06  	&      0.07    	&    	0.07  	&     	0.08 		\\
1.5\%: &C$_{4-5}$ &   	0.10   	&      5.73		&       0.06		&       0.08		&      9.01  	&      0.08    	&    	0.08   	&     	0.11 		\\
3.3\%:  &C$_{6-9}$ & 	0.42    	&  	 0.30  	&    	 0.30 	&    	 0.31 	&    	0.32   	&     	7.97    	&    	0.28   	&     	0.36		\\
3.3\%:  &C$_{10-14}$ & 0.58  	&      0.45 		&     	 0.43		&     	 0.44 	&    	0.44 		&      	0.41   	&     	7.25    	&    	0.44		\\
2.1\%: &C$_{15-19}$ & 0.42		&      0.30   	&  	 0.31  	&   	 0.33		&      	0.34 		&      0.34   	&     	0.32   	&	6.84	\\\hline

 \end{tabular*}
 \end{center}
 \end{table*}
 
\end{document}